\documentclass[journal=jpccck,manuscript=article,singlespacing]{achemso}
\usepackage[T1]{fontenc} 
\usepackage{amssymb}
\usepackage{graphicx}
\usepackage{subcaption}
\usepackage[outdir=./]{epstopdf}
\usepackage{bm}
\usepackage{color}
\usepackage{amsmath}
\usepackage{setspace}

\newcommand{\beq}{\begin{equation}}
\newcommand{\eeq}{\end{equation}}
\newcommand{\beqar}{\begin{eqnarray}}
\newcommand{\eeqar}{\end{eqnarray}}
\newcommand{\bea}{\begin{eqnarray}}
\newcommand{\eea}{\end{eqnarray}}
\newcommand{\bcen}{\begin{center}}
\newcommand{\ecen}{\end{center}}

\newcommand{\ket}[1]{\left| #1 \right>}

\newcommand{\Imag}{\mathrm{Im}}

\author{Amikam Levy}
\affiliation{Department of Chemistry,  University of California, Berkeley, Berkeley, California 94720, United States}
\alsoaffiliation{The Raymond and Beverly Sackler Center for Computational Molecular and Materials Science, Tel Aviv University, Tel Aviv, Israel 69978}
\altaffiliation{Both authors contributed equally to this work}
\email{amikam@berkeley.edu}

\author{Lyran Kidon}
\affiliation{Department of Chemistry,  University of California, Berkeley, Berkeley, California 94720, United States}
\altaffiliation{Both authors contributed equally to this work}
\email{lyran@berkeley.edu}

\author{Jakob {B\"atge}}
\affiliation{Institute of Physics, University of Freiburg, Hermann-Herder-Strasse 3, 79104 Freiburg, Germany}

\author{Junichi Okamoto}
\affiliation{Institute of Physics, University of Freiburg, Hermann-Herder-Strasse 3, 79104 Freiburg, Germany}

\author{Michael Thoss}
\affiliation{Institute of Physics, University of Freiburg, Hermann-Herder-Strasse 3, 79104 Freiburg, Germany}

\author{David T. Limmer}
\affiliation{Department of Chemistry,  University of California, Berkeley, Berkeley, California 94720, United States}
\alsoaffiliation{Kavli Energy NanoScience Institute, Berkeley, California 94720, United States}
\alsoaffiliation{Materials Sciences Division, Lawrence Berkeley National Laboratory, Berkeley, California 94720, United States}

\author{Eran Rabani}
\affiliation{Department of Chemistry,  University of California, Berkeley, Berkeley, California 94720, United States}
\alsoaffiliation{Materials Sciences Division, Lawrence Berkeley National Laboratory, Berkeley, California 94720, United States}
\alsoaffiliation{The Raymond and Beverly Sackler Center for Computational Molecular and Materials Science, Tel Aviv University, Tel Aviv, Israel 69978}
\email{eran.rabani@berkeley.edu}

\title{Absence of Coulomb Blockade in the \\ Anderson Impurity Model\\ at the Symmetric Point}

\begin{document}

\begin{abstract}
\singlespacing
In this work, we investigate the characteristics of the electric
current in the so-called symmetric Anderson impurity model. We study
the nonequilibrium model using two complementary approximate methods,
the perturbative quantum master equation approach to the reduced
density matrix, and a self-consistent equation of motion approach to
the nonequilibrium Green's function. We find that at a particular
symmetry point, an interacting Anderson impurity model recovers the
same steady-state current as an equivalent non-interacting model, akin
a two-band resonant level model. We show this in the Coulomb
  blockade regime for both high and low temperatures, where either the
  approximate master equation approach and the Green's function method
  provide accurate results for the current. We conclude that the
steady-state current in the symmetric Anderson model at this regime
does not encode characteristics of a many-body interacting system.
\end{abstract}
%
\section{Introduction}
\singlespacing
The Anderson impurity model~\cite{anderson1961} is a fundamental model
for studying strongly correlated open quantum systems that appears in
many physical situations, including coupled quantum dots in
semiconductor heterostructures~\cite{Kastner1998,hanson2007spins} and
in molecular- and nano-electronics.\cite{Nitzan2003,Heath2003}
It is one of the simplest models of interacting particles and exhibits
complex many-body phenomena, such as the Coulomb
blockade,\cite{Beenakker1991} pair tunneling,\cite{koch2006} and the
Kondo effect.\cite{kondo1964,schrieffer1966}  Such nonequilibrium
steady-state effects of interacting open quantum systems, continue to
present a grand challenge for theory. As such, the Anderson impurity
model is often used as a benchmark for developing approximate methods
to study many-body physics of interacting particles, in and out of
equilibrium.

In recent years, numerous approximate methods have been developed to
study transport through nanoscale interacting systems. Among these are
quantum master equation approaches and their
generalizations,\cite{Datta1990,Mukamel2006,Leijnse2008,Esposito2009,Esposito2010,dou2015,dorda2015}
approaches based on the nonequilibrium Green's function
formalism,\cite{hettler1998,Datta2000,Xue2002,Galperin07,galperin07b,Haug2008,galperin17,stefanucci_nonequilibrium_2013}
and quasi-classical mapping techniques.\cite{li2013,li2014} More
recently, numerically exact approaches (namely, methods that allow for
a systematic convergence of the results) have been proposed that allow
for an assessment of the approximate methods in certain regimes of
interactions and temperatures. Most notable are real-time path
integral methods based on diagrammatic expansions of the hybridization
or onsite
interactions,\cite{rabani2008,weiss_iterative_2008,schiro_real-time_2009,werner_diagrammatic_2009,gull10_bold_monte_carlo,Segal10,cohen_memory_2011,Hartle2013,Cohen-Gull-Reichman2015-introducing-inchworm}
renormalization group
techniques,\cite{schmitteckert_nonequilibrium_2004,anders_real-time_2005,bulla2008numerical}
or many-body wavefunction techniques.\cite{Wang2009} Benchmarks of the
various approximation schemes are often limited to the so-called
symmetric Anderson
model,\cite{werner_diagrammatic_2009,li2013,hofstetter1999,dickens2001,motahari2016}
where the empty and fully occupied states of the impurity are
degenerate.

In this work, we show that the steady-state current in the Anderson
impurity model at the symmetric point coincides exactly with an
equivalent noninteracting model, transition. The results presented in
this study are derived from two complementary approximate methods: 1)
the quantum master equation (QME) approach and 2) the equation of
motion (EOM) nonequilibrium Green's function (NEGF)
approach.\cite{Schwinger1961,Keldysh1964} The QME approach is
  adequate in the weak system-leads coupling limit and at high
  temperatures, for arbitrary onsite interactions, whereas the
  EOM-NEGF approach used here is accurate for small onsite
  interactions, but is not limited to weak system-bath
  couplings. These methods do not account for the Kondo effect or
  pair-tunneling. As such, this work focuses on the many-body physics
  of the Anderson model, which is manifested in the Coulomb blockade
  regime.

\section{Model and methods}       
\label{sec:model}
The Anderson impurity model is defined by the Hamiltonian
$H=H_{S}+H_{B}+V$, where \beqar H_{S}= &
\underset{\sigma=\uparrow,\downarrow}{\sum}
\varepsilon_{\sigma}d_{\sigma}^{\dagger}d_{\sigma}+Ud_{\uparrow}^{\dagger}d_{\uparrow}d_{\downarrow}^{\dagger}d_{\downarrow}
\eeqar describes the impurity (or dot), referred to simply as the
`system Hamiltonian', \beqar H_{B}= & \underset{\underset{k\in \rm
    L,\rm R}{
    \sigma=\uparrow,\downarrow}}{\sum}\varepsilon_{k}c_{k\sigma}^{\dagger}c_{k\sigma}
\eeqar describes the noninteracting fermionic baths (or leads), and
\beqar V= & \underset{\underset{k\in \rm L,\rm R}{
    \sigma=\uparrow,\downarrow}}{\sum}
t_{k}d_{\sigma}^{\dagger}c_{k\sigma}+{\rm h.c.}, \eeqar describes the
hybridization between the system and the leads. In the above,
$d_{\sigma}^{\dagger}\ \left(d_{\sigma}\right)$ are the creation
(annihilation) operators of an electron on the dot with spin
$\sigma=\uparrow,\downarrow$ with one-body energy of
$\varepsilon_{\sigma}$, $U$ is the onsite Hubbard interaction,
$c_{k\sigma}^{\dagger}\ \left(c_{k\sigma}\right)$ are the creation
(annihilation) operators of an electron in mode $k$ of the leads with
energy $\varepsilon_{k}$, and $t_{k}$ is the hybridization between the
dot and mode $k$ in the lead. The coupling to the quasi-continuous
leads is modeled by the leads' spectral function,
$\Gamma_{\ell}\left(\omega\right)$, where
$t_{k\in\ell}=\sqrt{\Gamma_{\ell}\left(\ensuremath{\varepsilon}_{k}\right)\Delta\omega/2\pi}$
is the coupling between the dot and the $k$-th mode of the $\ell=\rm
L,R$ bath, and $\Delta\omega$ is the discretization of the leads
energy spectrum. Throughout, we take Planck's constant to be 1, and
assume that the coupling to the leads is spin-independent.  When a
wide-band limit is assumed,
$\Gamma_\ell\left(\omega\right)=\Gamma_\ell$ is taken to be
independent of $\omega$.  \par
\begin{figure}[t]
\center{\includegraphics[width=8.25cm]{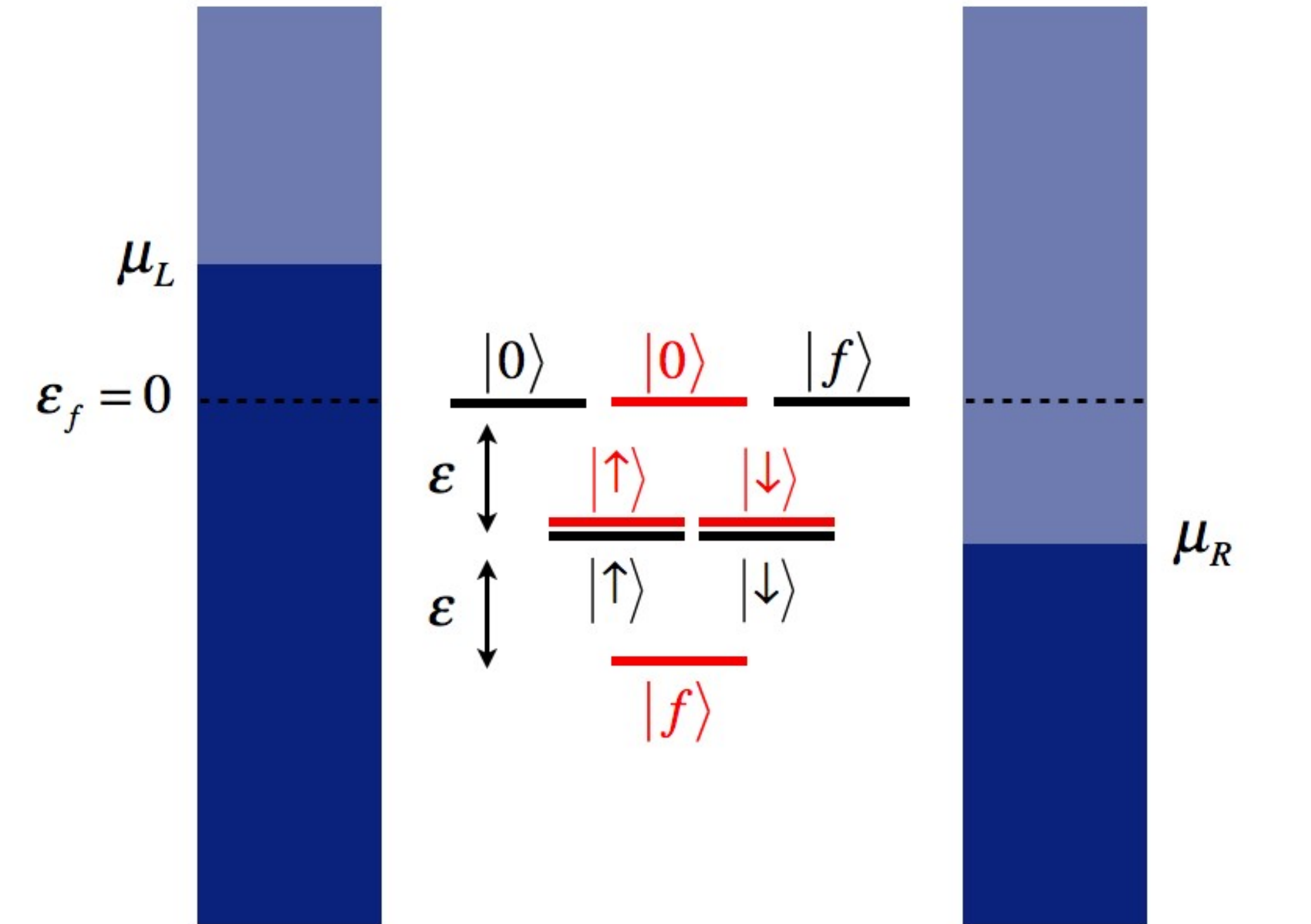}}
\caption{Schematic representation of the many-body energy levels of
  the system Hamiltonian, $H_{S}$. In red: the non-interacting system
  $U=0$, in black: the symmetric model $U=-2\varepsilon$. The Fermi
  energy of the leads, $\varepsilon_{f}$, around which the chemical
  potentials are biased, is set to zero.}
\label{fig:schema}
\end{figure}

 We will focus on two regimes of the
model, the non-interacting regime where $U=0$ and
$\varepsilon_{\uparrow}=\varepsilon_{\downarrow}=\varepsilon$, and the
symmetric point where
$\varepsilon_{\uparrow}=\varepsilon_{\downarrow}=-U/2$. The system is
held out of equilibrium by an electric bias by setting the leads'
chemical potentials to be $\mu_{\ell}=\varepsilon_f+V_\ell$, where
$\varepsilon_f$ is the Fermi energy of the leads and $V_\ell$ is the
bias on the $\ell$th lead. A symmetric bias is said to be applied when
$\varepsilon_f=0$.  The many-body energy levels in both regimes of the
system Hamiltonian, $H_{S}$, are schematically introduced in
Fig.~\ref{fig:schema}. An empty dot, $\ket{0}$ with energy $0$, a
singly-occupied dot with an electron in either spin,
$\ket{\sigma}=d^{\dagger}_{\sigma}\ket{0}$ with energy $\varepsilon$,
and a doubly occupied (or full) dot
$\ket{f}=d^{\dagger}_{\uparrow}d^{\dagger}_{\downarrow}\ket{0}$, with
energy $2\varepsilon+U$.  We may note that in both cases studied all
one-body transitions are given by the same energy difference, which is
equal to $\varepsilon$. This is a key feature of the symmetric
Anderson model, giving rise to identical one-body transition
probabilities as the noninteracting case.

\par We analyze the model using two methods, a quantum master equation
approach and a nonequilibrium Green's function approach. Although NEGF
is typically more general and has a wider regime of validity than the
QME approach, we present both results, as the perturbative QME
approach facilitates a simple derivation. In the supporting material
we discuss the validity of the approximations in more detail and
compare the two methods in their appropriate regimes.
\subsection{Quantum master equation}
The quantum master equation description~\cite{breuer,nitzan06} assumes
weak-coupling between the dot and the leads, and is derived using
second order perturbation theory in the dot-lead coupling
strength. The time scale at which correlations in the leads decay
should be smaller than the time scale for the leads to induce a
significant change in the dot.  This implies that the QME is valid
when the temperature of the leads is sufficiently high with respect to
the coupling strength (see supporting information for details). It is
further assumed that there are no initial correlations between the dot
and the leads, and that the leads are in local thermal
equilibrium. Since the system Hamiltonian is diagonal in the many-body
basis and the dot-lead coupling is bi-linear, the populations are
coupled directly and not through the coherences. In other words, the
equations of motion for the populations (diagonal terms of the reduced
density matrix) and the coherences (non-diagonal terms) are decoupled
in the many-body basis. Thus, if the initial state of the system is
diagonal in the dot-energy eigen-basis or if we are interested in
physical quantities that depend on populations alone, then the QME is
reduced to a rate equation for the many-body states:
\beq
\left(\begin{array}{c}
\dot{p}_{0}\\
\dot{p}_{\uparrow}\\
\dot{p}_{\downarrow}\\
\dot{p}_{f}
\end{array}\right)=M\left(\begin{array}{c}
p_{0}\\
p_{\uparrow}\\
p_{\downarrow}\\
p_{f}
\end{array}\right),
\eeq
Here $p_0$ is the probability of occupying the empty dot state $\ket{0}$ , $p_{\uparrow}$ and $p_{\downarrow}$ are the probabilities of occupying the states $\ket{\uparrow}$ and $\ket{\downarrow}$ respectively, and $p_f$ is the probability of having a fully occupied dot corresponding to the state $\ket{f}$.     
The transition matrix is given by 
\beq
M = \sum_{\ell=\rm L,R}\left(\begin{array}{cccc}
-\left(\gamma^\ell_{\uparrow0}+\gamma^\ell_{\downarrow0}\right) & \gamma^\ell_{0\uparrow} & \gamma^\ell_{0\downarrow} & 0\\
\gamma^\ell_{\uparrow0} & -\left(\gamma^\ell_{0\uparrow}+\gamma^\ell_{f\uparrow}\right) & 0 & \gamma^\ell_{\uparrow f}\\
\gamma^\ell_{\downarrow0} & 0 & -\left(\gamma^\ell_{0\downarrow}+\gamma^\ell_{f\downarrow}\right) & \gamma_{\downarrow f}\\
0 & \gamma^\ell_{f\uparrow}& \gamma^\ell_{f\downarrow} & -\left(\gamma^\ell_{\uparrow f}+\gamma^\ell_{\downarrow f}\right)
\end{array}\right),
\eeq
where $\gamma^\ell_{ij}$ is the transition rate from state $j$ to state $i$ induced by the collective $\ell$ bath.
The above rates can be calculated using Fermi's golden-rule and satisfy local detailed balance such that 
$\gamma^\ell_{ij}=\gamma^\ell_{ji}\exp [\beta_\ell (\varepsilon_j-\varepsilon_i)]$ where $\beta_\ell$ is the inverse temperature of the $\ell$-th lead and $\varepsilon_i$ is the energy of the eigenstate $i$. 

Assuming symmetry between spin up and spin down, $\varepsilon_{\uparrow}=\varepsilon_{\downarrow}\equiv \varepsilon$, in the wide-band approximation, we can explicitly write

\beqar
%
\gamma^\ell_{\uparrow0}&=&\gamma^\ell_{\downarrow0} \equiv \Gamma_\ell f^\ell(\varepsilon-\mu_\ell) \\ \nonumber
\gamma^\ell_{0\uparrow}&=&\gamma^\ell_{0\downarrow} \equiv \Gamma_\ell f^\ell(-\varepsilon+\mu_\ell) \\ \nonumber
\gamma^\ell_{f\uparrow}&=&\gamma^\ell_{f\downarrow}\equiv \Gamma_\ell f^\ell(\varepsilon+U-\mu_\ell)\\ \nonumber
\gamma^\ell_{\uparrow f}&=&\gamma^\ell_{\downarrow f}\equiv \Gamma_\ell f^\ell(-\varepsilon-U+\mu_\ell),
\eeqar 
where $f^\ell(\varepsilon-\mu_\ell)=\left( 1 + \exp [\beta_\ell (\varepsilon-\mu_\ell)] \right)^{-1}$ is the Fermi-distribution with the chemical potential $\mu_\ell$.
We note that the results presented in the manuscript are not limited to a wide-band approximation and are valid for any even spectral function, $\Gamma_\ell(\varepsilon)=\Gamma_\ell(-\varepsilon)$.

The current from the $\ell$ bath is calculated according to
\begin{equation}
\label{eq:current}
I_{\ell}(t)=\frac{e}{2}\sum_{ij}\left(N_{i}-N_{j}\right)J_{ij}^{\ell} \quad \text{where} \quad J_{ij}^{\ell}=\gamma_{ij}^{\ell}p_{j}(t)-\gamma_{ji}^{\ell}p_{i}(t),
\end{equation}
where $N_{i}$ is the number of electrons in state $i$, and $J_{ij}^{\ell}$ is the probability current from state $j$ to state $i$, due to the coupling to the $\ell$ bath, and $e$ is the charge of the electron.
%
\subsection{Equation of motion nonequilibrium Green's functions}
To support our findings we further investigate the Anderson model
  at the symmetric point using an EOM approach to
  NEGF.\cite{cuniberti2007,galperin07b,Scheck2013,Levy2013,Levy2013a}
  We note that the approach taken here is valid to all orders of
  $\Gamma$ but is exact to order $\Gamma^2U$. Thus, in the limit $U
  \rightarrow 0$ the NEGF approach recovers the exact noninteracting
  results for any system-bath coupling strength. The approach does not
  capture the Kondo and pair-tunneling effects but is capable of
  describing the Coulomb blockade
  regime.\cite{cuniberti2007,galperin07b,Scheck2013,Levy2013,Levy2013a}
We begin by defining the impurity one-body nonequilibrium contour
ordered Green's
function,\cite{Schwinger1961,Keldysh1964,Stefanucci2006}
\begin{equation}
G_{\sigma}\left(\tau,\tau'\right)=-i\left\langle \hat{T}_{c}\left[d_{\sigma}\left(\tau\right)d_{\sigma}^{\dagger}\left(\tau'\right)\right]\right\rangle 
\end{equation}
where $\hat{T}_{c}$ is the contour ordering operator and the average
$\langle \cdots \rangle$ should be interpreted as the trace over the
many particle Hilbert space with the equilibrium density matrix at
$t=-\infty$. The equation of motion for the Green's functions on the
Keldysh contour is obtained from the Heisenberg equation for the time
propagation of the operators, leading to an equation of motion which
introduces new, higher-order Green's functions, for which equations of
motion are also derived. For interacting Hamiltonians, this method
produces a hierarchy of inter-dependent equations for Green's
functions of higher orders. A common procedure is to truncate the
equations by choosing an appropriate approximation (or closure), after
which the equations are solved self-consistently. In this work we use
a closure that is exact for the noninteracting limit, where $U=0$, for
any value of the system-leads coupling $\Gamma$, and at any
temperature, reproducing the exact resonant level model solution. It
also provides a good approximation at sufficiently large values of
$U$, in the Coulomb blockade regime.\cite{cuniberti2007,Haug2008} A
detailed derivation of the closure is provided in the Supplementary
Information.  \par We solve the equations of motion on the Keldysh
contour and analytically continued to the real time axis using
Langreth rules.\cite{Langreth76} The integral equations for the
retarded and lesser NEGFs in steady state are transformed to the
frequency domain, resulting in the following algebraic equations:
\begin{align}
G_{\sigma}^{r}\left(\omega\right)= & g_{\sigma}^{\left(0\right)r}\left(\omega\right)+g_{\sigma}^{\left(0\right)r}\left(\omega\right)\Sigma_{\sigma}^{r}\left(\omega\right)G_{\sigma}^{r}\left(\omega\right)+Ug_{\sigma}^{\left(0\right)r}\left(\omega\right)G_{\sigma}^{\left(2\right)r}\left(\omega\right),\\
G_{\sigma}^{\left(2\right)r}\left(\omega\right)= & g_{\sigma}^{\left(U\right)r}\left(\omega\right)\left\langle n_{\bar{\sigma}}\right\rangle +g_{\sigma}^{\left(U\right)r}\left(\omega\right)\Sigma_{\sigma}^{r}\left(\omega\right)G_{\sigma}^{\left(2\right)r}\left(\omega\right),\nonumber \\
G_{\sigma}^{<}\left(\omega\right)= & g_{\sigma}^{\left(0\right)<}\left(\omega\right)+g_{\sigma}^{\left(0\right)r}\left(\omega\right)\Sigma_{\sigma}^{r}\left(\omega\right)G_{\sigma}^{<}\left(\omega\right)+g_{\sigma}^{\left(0\right)r}\left(\omega\right)\Sigma_{\sigma}^{<}\left(\omega\right)G_{\sigma}^{a}\left(\omega\right)\nonumber \\
  +& g_{\sigma}^{\left(0\right)<}\left(\omega\right)\Sigma_{\sigma}^{a}\left(\omega\right)G_{\sigma}^{a}\left(\omega\right)+Ug_{\sigma}^{\left(0\right)r}\left(\omega\right)G_{\sigma}^{\left(2\right)<}\left(\omega\right)+Ug_{\sigma}^{\left(0\right)<}\left(\omega\right)G_{\sigma}^{\left(2\right)a}\left(\omega\right),\nonumber \\
G_{\sigma}^{\left(2\right)<}\left(\omega\right)= & g_{\sigma}^{\left(U\right)<}\left(\omega\right)\left\langle n_{\bar{\sigma}}\right\rangle +g_{\sigma}^{\left(U\right)r}\left(\omega\right)\Sigma_{\sigma}^{r}\left(\omega\right)G_{\sigma}^{\left(2\right)<}\left(\omega\right)\nonumber \\
  +& g_{\sigma}^{\left(U\right)r}\left(\omega\right)\Sigma_{\sigma}^{<}\left(\omega\right)G_{\sigma}^{\left(2\right)a}\left(\omega\right)+g_{\sigma}^{\left(U\right)<}\left(\omega\right)\Sigma_{\sigma}^{a}\left(\omega\right)G_{\sigma}^{\left(2\right)a}\left(\omega\right),\nonumber 
\end{align}
where $\bar{\sigma}$ is the opposite spin to
$\sigma$,
\begin{align}
G_{\sigma}^{r}\left(t,t'\right) &= -i\theta\left(t-t'\right)\left\langle \left\{ d_{\sigma}\left(t\right),d_{\sigma}^{\dagger}\left(t'\right)\right\} \right\rangle ,\\
G_{\sigma}^{<}\left(t,t'\right) &= i\left\langle d_{\sigma}^{\dagger}\left(t'\right)d_{\sigma}\left(t\right)\right\rangle ,\nonumber \\
G_{\sigma}^{\left(2\right)r}\left(t,t'\right) &= -i\theta\left(t-t'\right)\left\langle \left\{ d_{\sigma}\left(t\right)n_{\bar{\sigma}}\left(t\right),d_{\sigma}^{\dagger}\left(t'\right)\right\} \right\rangle ,\nonumber \\
G_{\sigma}^{\left(2\right)<}\left(t,t'\right) &= i\left\langle d_{\sigma}^{\dagger}\left(t'\right)d_{\sigma}\left(t\right)n_{\bar{\sigma}}\left(t\right)\right\rangle ,\nonumber 
\end{align}
and $G\left(\omega\right)=\int_{-\infty}^{\infty}d\tau\,G\left(\tau\right)e^{i\omega\tau}$, with 
$\tau=t-t'$. In the above, the steady state population of spin $\sigma$
is given by
\begin{equation}
\left\langle n_{\sigma}\right\rangle =\int_{-\infty}^{\infty}\frac{d\omega}{2\pi}G_{\sigma}^{<}\left(\omega\right),
\end{equation}
and we used the following definitions for the noninteracting Green's functions:
\begin{align}
g_{\sigma}^{\left(\delta\right)r}\left(\omega\right) & =\frac{1}{\omega-\varepsilon_{\sigma}-\delta+i\eta},\\
g_{\sigma}^{\left(\delta\right)<}\left(\omega\right) & =2\pi\left\langle d^{\dagger}d\right\rangle _{0}\delta\left(\omega-\varepsilon_{\sigma}-\delta\right),\nonumber 
\end{align}
with $\delta=0$ or $\delta=U$. The self energy due to the coupling to the leads are given by:\cite{meir_landauer_1992}
\begin{align}
\Sigma_{\sigma}^{r}\left(\omega\right) & = \Sigma_{\rm L}^{r}\left(\omega\right)+\Sigma_{\rm R}^{r}\left(\omega\right) = -\frac{i}{2}\left(\Gamma_{\rm L}\left(\omega\right)+\Gamma_{\rm R}\left(\ensuremath{\omega}\right)\right),\\
\Sigma_{\sigma}^{<}\left(\omega\right) & = \Sigma_{L}^{<}\left(\omega\right)+\Sigma_{\rm R}^{<}\left(\omega\right) = i\left(\Gamma_{\rm L}\left(\omega\right)f^{\rm L}\left(\omega-\mu_{\rm L}\right)+\Gamma_{\rm R}\left(\omega\right)f^{\rm R}\left(\omega-\mu_{\rm R}\right)\right).\nonumber 
\end{align}
The above self energy is identical to the resonant level model self
energy, where again $f^\ell(\omega)$ is the Fermi distribution with inverse temperature $\beta_\ell$ and a symmetric spectral function was assumed,
$\Gamma_{\ell}\left(\omega\right)=\Gamma_{\ell}\left(-\omega\right)$. 
The current at steady-state due to the $\ell$-th bath is given by, \cite{meir_landauer_1992}
\begin{equation}
I_{\ell}=2e \sum_{\sigma}\int\frac{d\omega}{2\pi}\left[i\Sigma_{\ell}^{<}\left(\omega\right)\Imag\left\{ G^{r}_{\sigma}\left(\omega\right)\right\} +\Sigma_{\ell}^{r}\left(\omega\right)G^{<}_{\sigma}\left(\omega\right)\right].
\end{equation}
\section{Results and discussion}
\label{sec:result}
\begin{figure}[!htb]
\minipage{0.33\textwidth}
  \includegraphics[width=\linewidth]{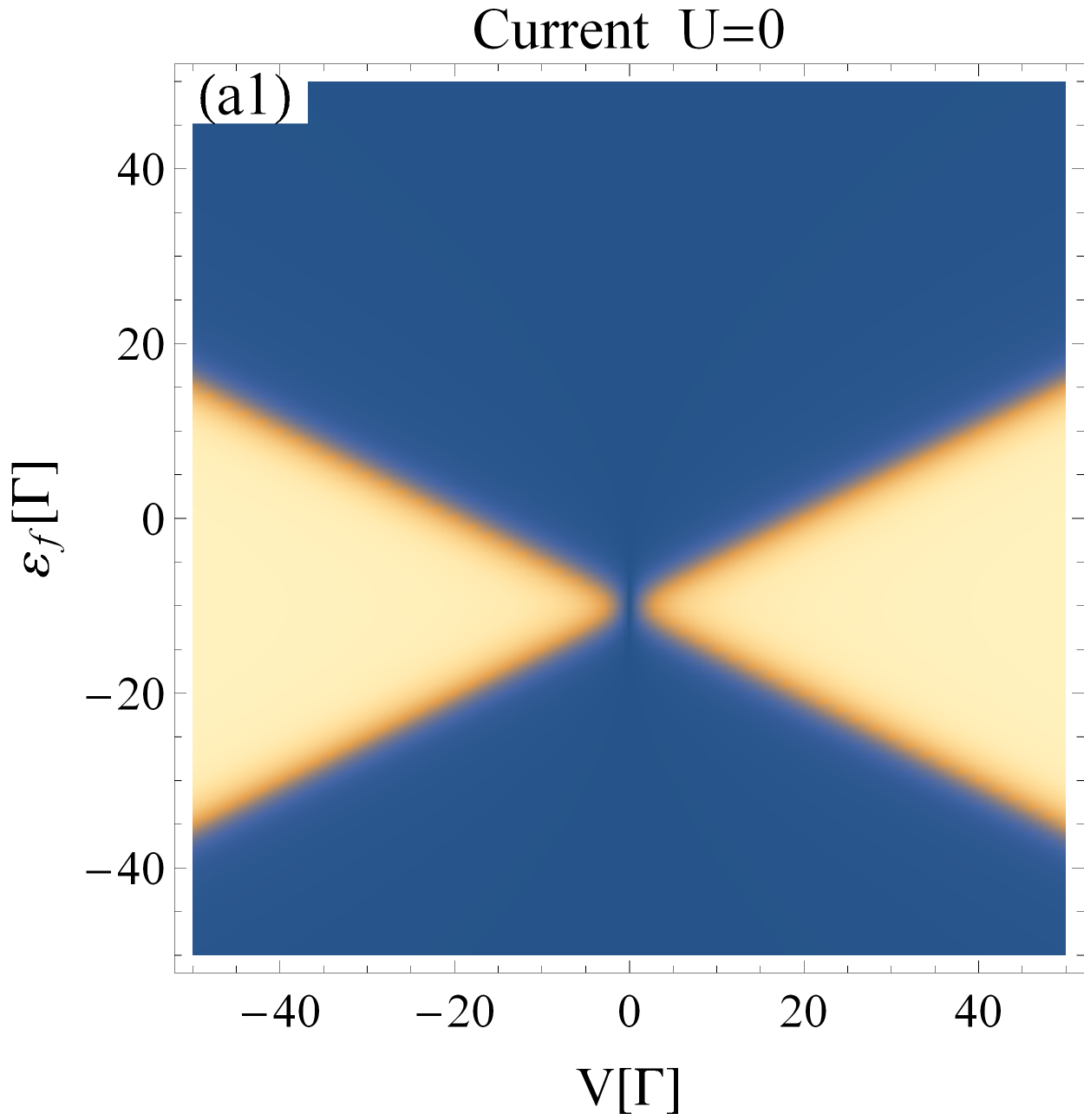}
\endminipage\hfill
\minipage{0.3\textwidth}
  \includegraphics[width=\linewidth]{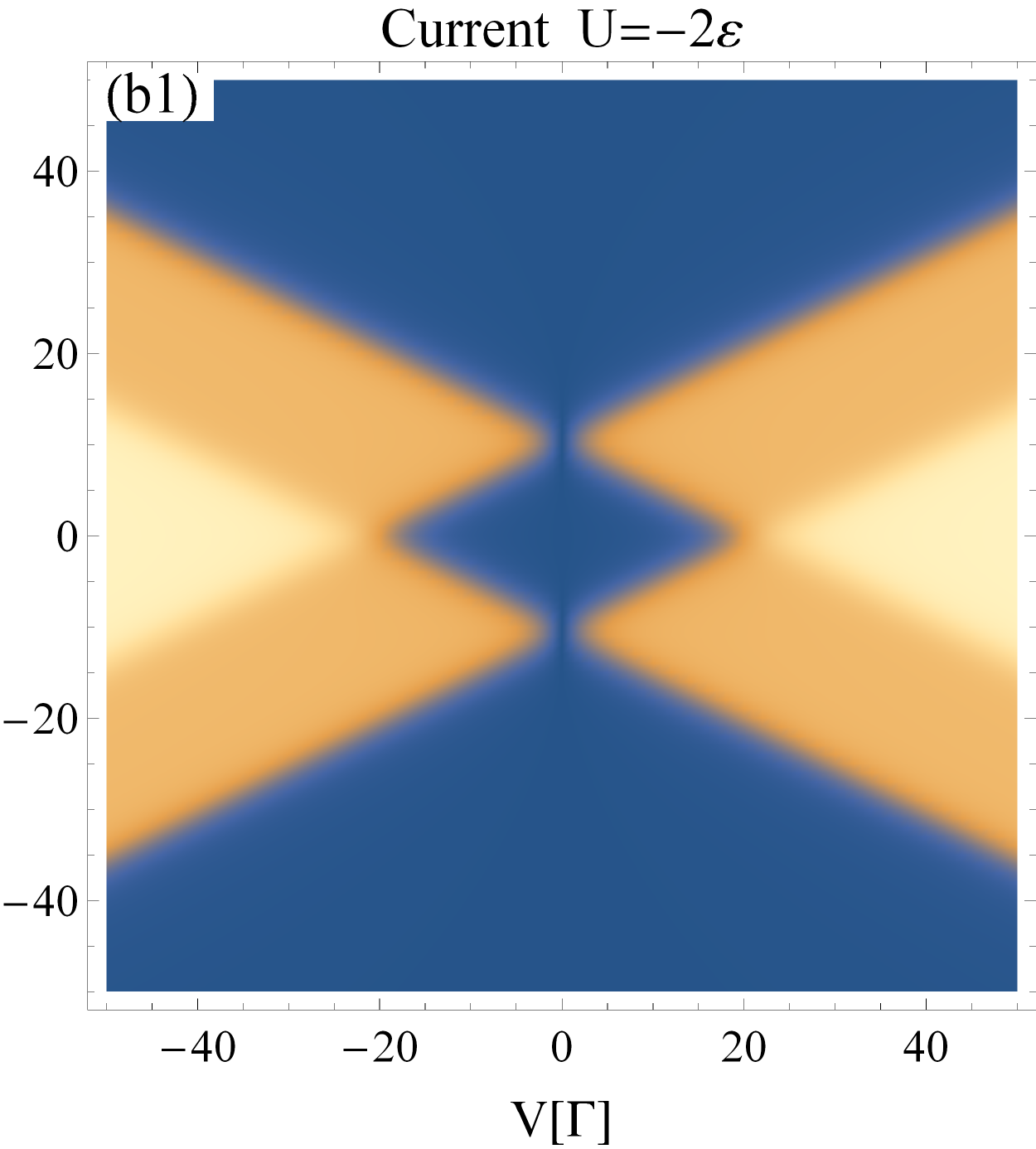}
\endminipage\hfill
\minipage{0.35\textwidth}%
  \includegraphics[width=\linewidth]{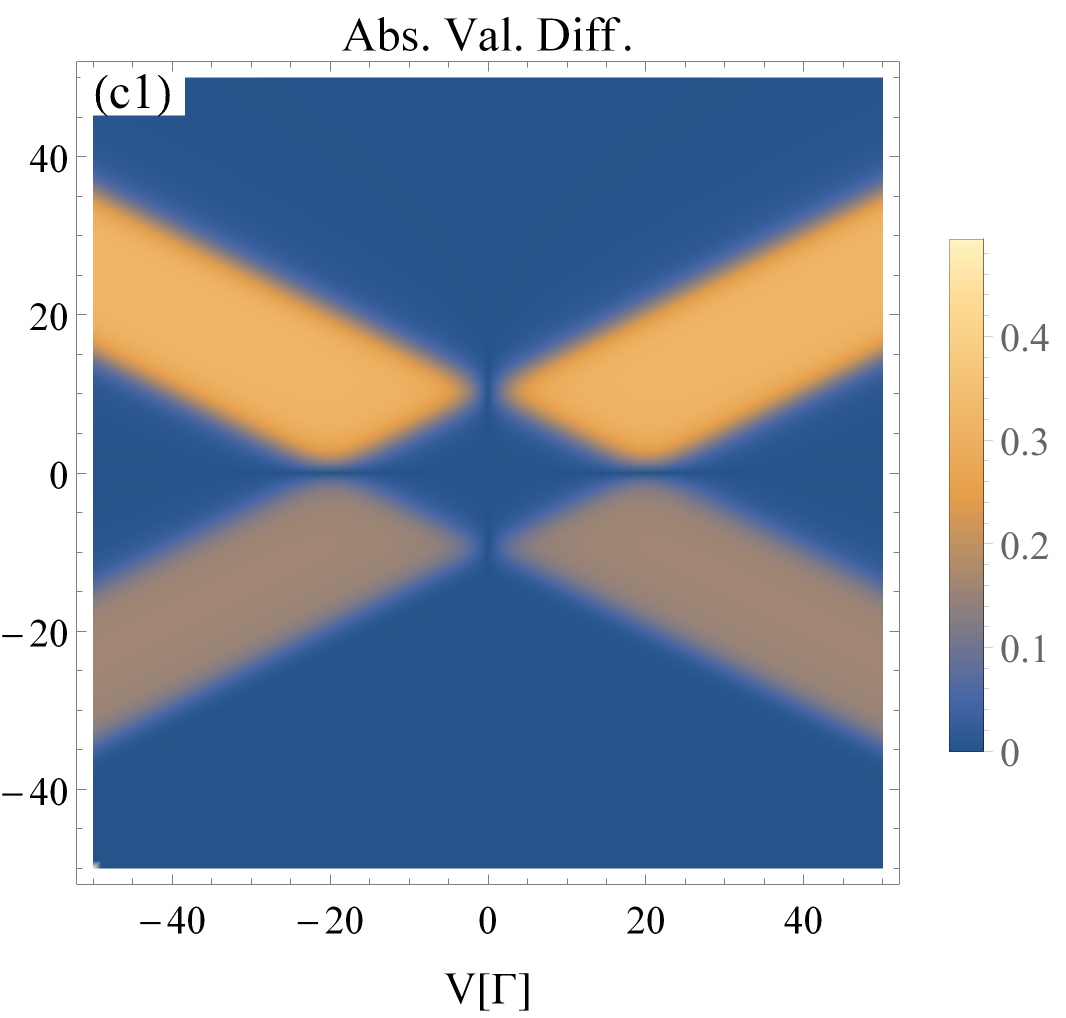}
\endminipage

\minipage{0.33\textwidth}
  \includegraphics[width=\linewidth]{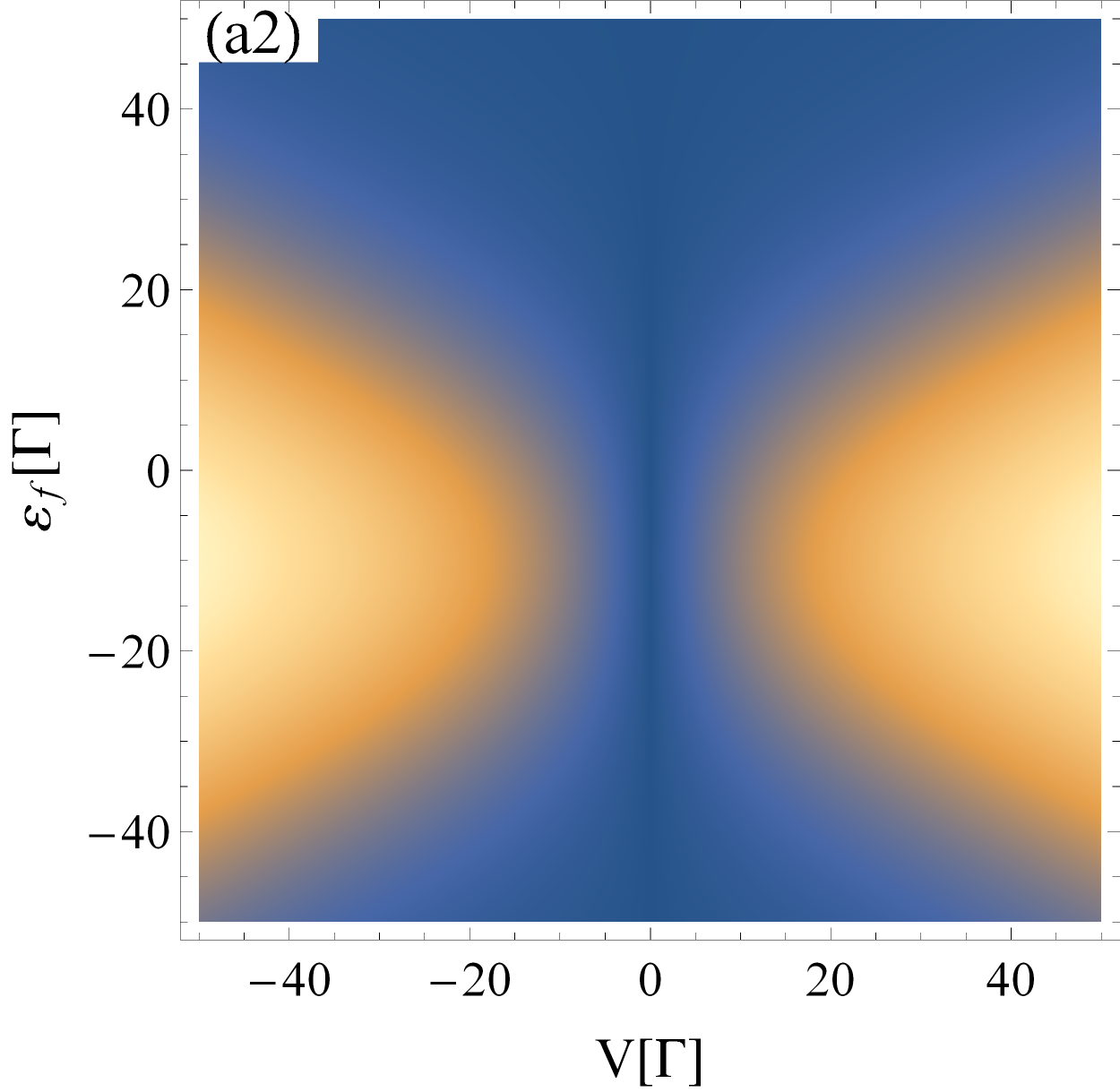}
\endminipage\hfill
\minipage{0.3\textwidth}
  \includegraphics[width=\linewidth]{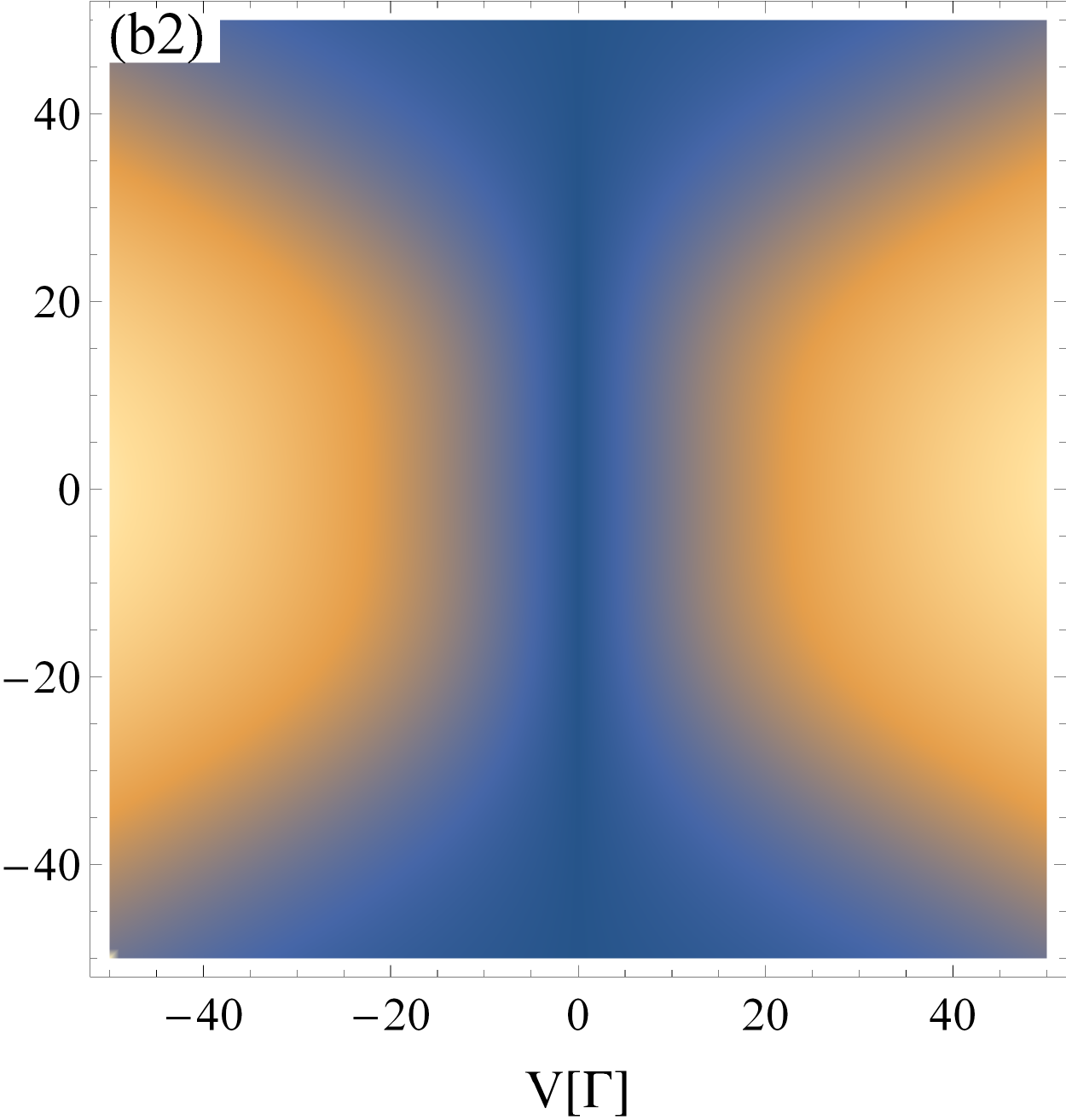}
\endminipage\hfill
\minipage{0.35\textwidth}%
  \includegraphics[width=\linewidth]{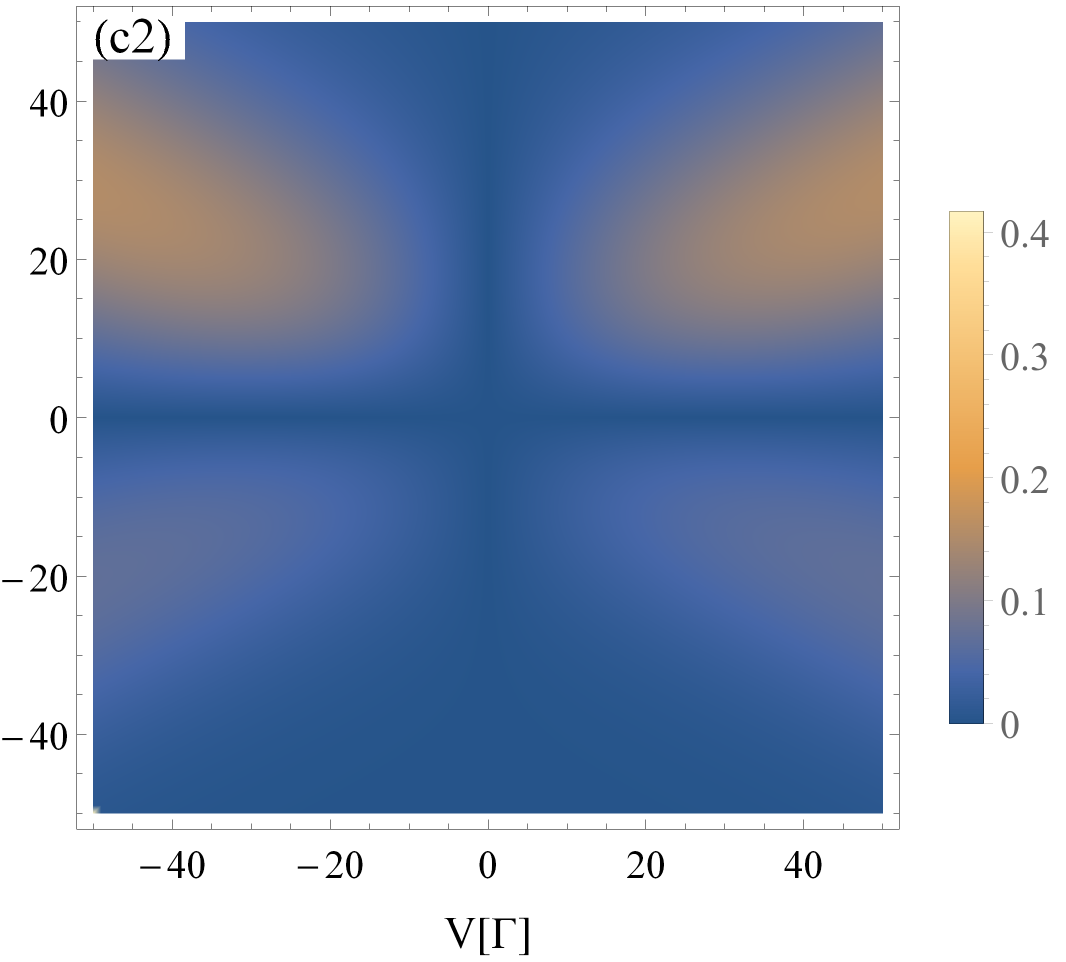}
\endminipage
\caption{The magnitude of the steady state current as a function of the Fermi-energy $\varepsilon_f$ and the bias voltage between the leads $V$. Panels (a1) and (a2) are for the noninteracting system, $U=0$. Panels (b1) and (b2) are for the interacting system with $U=-2\varepsilon$. Panels (c1) and (c2) display the absolute value of the difference between the interacting and noninteracting results.
Upper and lower panels were computed for $T_{\rm L}=T_{\rm R}=\Gamma$ and $T_{\rm L}=T_{\rm R}=10\Gamma$, respectively.  Other parameters used: $\Gamma_{\rm L}=\Gamma_{\rm R}=\Gamma/2$, and $\varepsilon=-10\Gamma$. 
} 
\label{fig:fig1}
\end{figure} 


In Fig.~\ref{fig:fig1} we plot the magnitude of the current for the noninteracting (left panels) and interacting (middle panels) models, $I_{{\rm L}}^{(0)}$ and $I_{ {\rm L}}^{(U)}$, respectively. The results are plotted as a function of  the Fermi energy ($\varepsilon_f$) and the bias voltage applied between the leads ($V=\mu_L-\mu_R$), with the chemical potentials of the leads set to $\mu_{_{\rm L,R}}=\varepsilon_f\pm V/2$. Two different temperatures were considered: $T_{\rm L}=T_{\rm R}=\Gamma$ (upper panels) and $T_{\rm L}=T_{\rm R}=10\Gamma$ (lower panels). To directly compare the results between the interacting and noninteracting models, we also plot the absolute value of the difference between the two (right panels). Results are shown for the EOM-NEGF approach, but a similar qualitative picture emerges within the QME formalism.

Focusing on the results for the noninteracting system (Fig.~\ref{fig:fig1}, panels (a1) and (a2)), as expected, we find significant values for the current at a finite bias when $\varepsilon_f - \varepsilon < |V|/2$, leading to the well-known diamond-like current characteristics, broadened by the temperature $T$. The picture is a bit more evolved for the interacting case (Fig.~\ref{fig:fig1}, panels (b1) and (b2)), where a double diamond-like current characteristic shape is observed for $U=-2\varepsilon$. The two-step value of the current results from the well-known Coulomb blockade, where the bias voltage is not sufficiently large to overcome the onsite repulsion, and only one conductance channel is open at intermediate bias voltage values. Only when $|V|$ becomes sufficiently large compared to $U$ an additional conducting channel opens up, and consequently the current increases to its maximal value. We note that the Coulomb blockade phenomenon is suppressed in the high temperature limit, as would be expected, where the current rises gradually to its maximum value.

\begin{figure}[!htb]
\minipage{0.45\textwidth}
  \includegraphics[width=\linewidth]{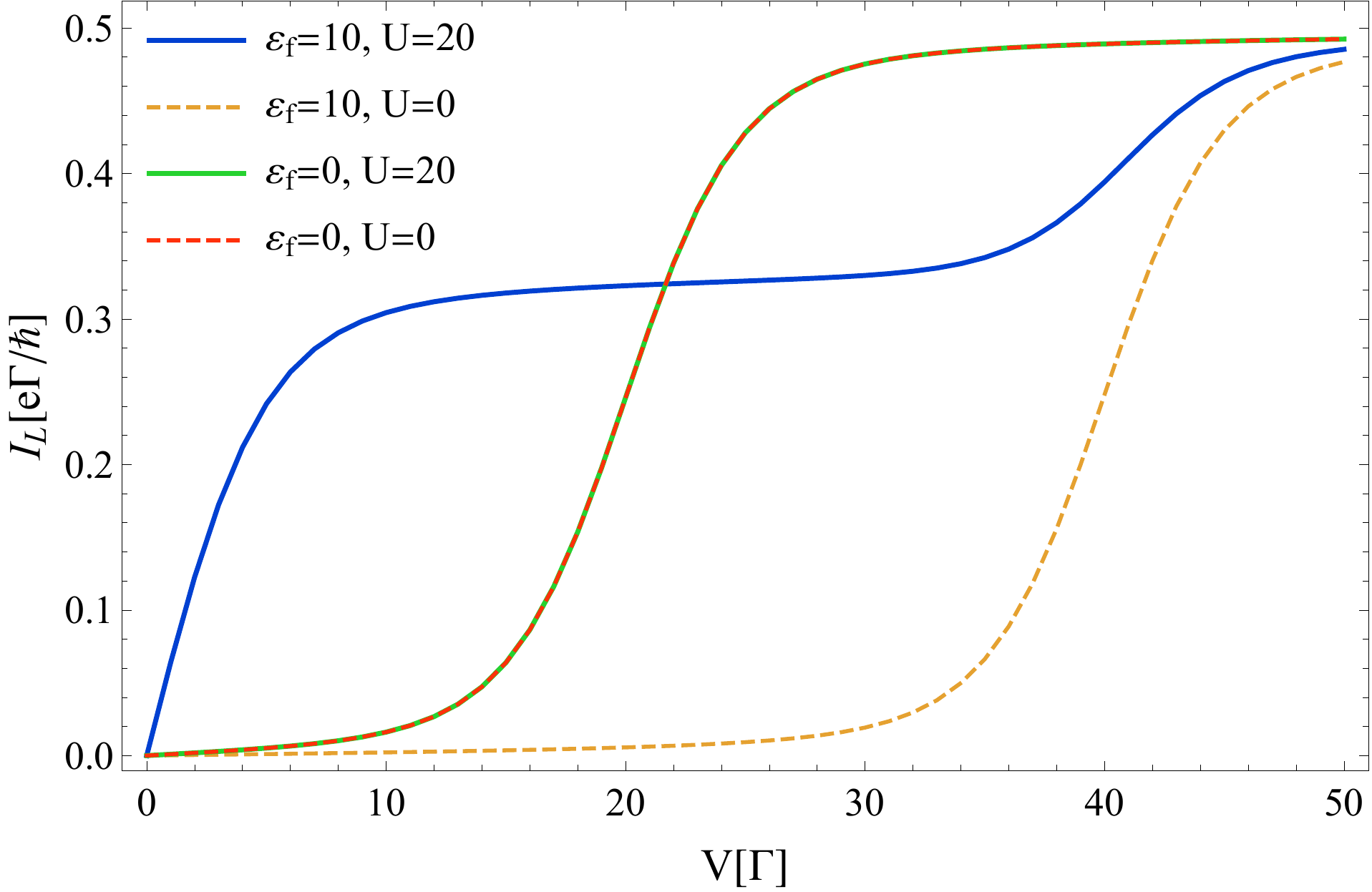}
\endminipage\hfill
\minipage{0.45\textwidth}
  \includegraphics[width=\linewidth]{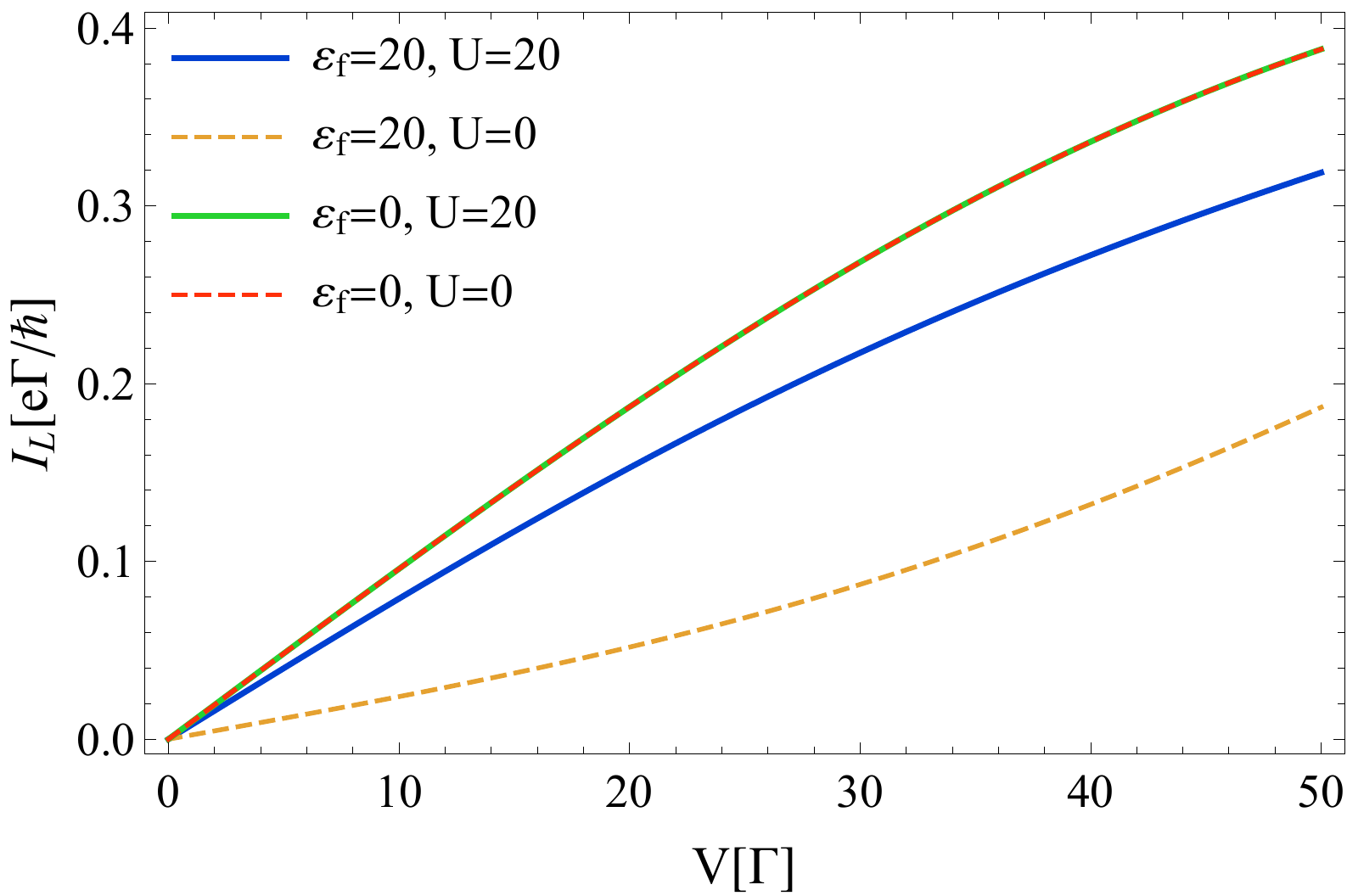}
\endminipage\
\caption{Magnitude of the steady state current as a function of bias voltage at and away from the symmetric point for a) the low temperature conditions in Fig.2 (a1) and (b1) and b) the high temperature conditions in Fig.2 (a2) and (b2). The solid lines denote the Anderson model, whereas the dashed lines denote the noninteracting model.
}
\label{fig:fignew}
\end{figure}
\par

\begin{figure}[t]
\center{\includegraphics[width=8.25cm]{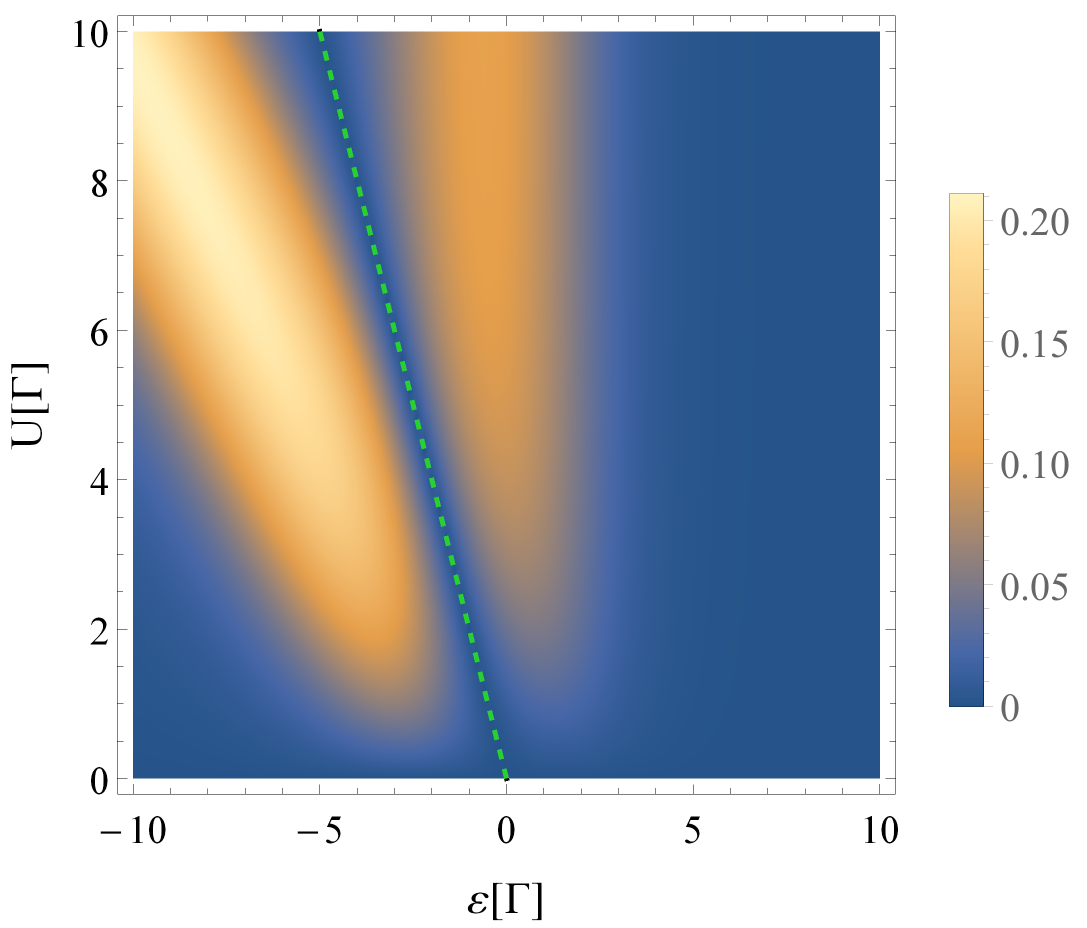}}
\caption{The absolute value of the difference between the steady-state currents of the interacting and noninteracting models at $\varepsilon_f=0$ and $V=4\Gamma$ for a range of gate voltages, $\varepsilon$ and onsite Hubbard interaction strength, $U$. The green dashed line represents the symmetric line defined by $U=-2\varepsilon$, where the difference between the two currents vanishes. Other parameters used: $\Gamma_{\rm L}=\Gamma_{\rm R}=\Gamma/2$, and $T_{\rm L}=T_{\rm R}=\Gamma$.}
\label{fig:fig2}
\end{figure}

Interestingly, when the bias voltage is applied symmetrically, i.e. $\varepsilon_f=0$, we find that the current-voltage characteristics are identical for both interacting and noninteracting models, suggesting that the Coulomb blockade is completely suppressed. This is clearly depicted in panels (c1) and (c2) of Fig.~\ref{fig:fig1}, where we plot the difference between the interacting and noninteracting currents, which diminishes as we approach the line $\varepsilon_f=0$ for all values of $V$. 
This is clarified in Fig.~\ref{fig:fignew} where we show cuts through the 2D plots of Fig.~\ref{fig:fig1} for two values of $\varepsilon_f$. When $\varepsilon_f \ne 0$, the I-V curves differ at both intermediate and high temperatures, while for a symmetric bias, $\varepsilon_f=0$, the interacting and noninteracting results overlap irrespective of the value of $T$. The results for $\varepsilon_f=0$ for different values of $U$ and $\varepsilon$ are summarized in Fig.~\ref{fig:fig2}, where we plot the difference between the interacting and noninteracting currents for a fixed bias voltage of $V=4\Gamma$. The green dashed line indicates the symmetric line where $U=-2\varepsilon$, for which the difference between the currents vanishes.

We now turn to rationalize this result. We begin by using the QME approach, which provides a simple relation between the interacting and noninteracting currents at $\varepsilon_f=0$. In the case where $\mu_{\rm L}=-\mu_{\rm R}\equiv\mu$ the master equation has a rather simple exact solution for the steady-state populations and currents (see supporting information). The steady-state current is given by
\beq
\label{eq:noninteracting_current}
I_{\rm L}^{\left(U=0\right)} =\frac{2e \Gamma_{\rm L} \Gamma_{\rm R} }{\Gamma_{\rm L} + \Gamma_{\rm R}}\left(f^{\rm L}(\varepsilon-\mu)-f^{\rm R}(\varepsilon+\mu) \right),
\eeq 
for the noninteracting case, whereas for the symmetric interacting model, the current takes a slightly more complicated form:
\beq
\label{eq:interacting_current}
I_{\rm L}^{\left(U=-2\varepsilon\right)}=\frac{2e
  \Gamma_L\Gamma_R\left(f^{\rm L}(\varepsilon-\mu)f^{\rm
    R}(\varepsilon-\mu)-f^{\rm L}(\varepsilon+\mu)f^{\rm
    R}(\varepsilon+\mu) \right)}{\Gamma_{\rm L} \left( f^{\rm
    L}(\varepsilon-\mu )+f^{\rm L}(\varepsilon+\mu)\right)+\Gamma_{\rm
    R} \left( f^{\rm R}(\varepsilon-\mu
  )+f^R(\varepsilon+\mu)\right)}.  \eeq When the temperatures of both
leads are balanced, i.e. $T_{\rm L}=T_{\rm R}$,
Eqs.~(\ref{eq:noninteracting_current}) and
(\ref{eq:interacting_current}) coincide exactly. Note that the leads'
spectral functions, $\Gamma_{\rm L}$ and $\Gamma_{\rm R}$, need not be
equal and the two expressions coincide as long as $\Gamma_L$ and
$\Gamma_R$ are even functions in energy.  In the case where
$\Gamma_{\rm L}=\Gamma_{\rm R}$, we obtain half-filling of the dot,
i.e.,
$n_{\uparrow}\big\vert_{U=-2\varepsilon}=n_{\downarrow}\big\vert_{U=-2\varepsilon}=\frac{1}{2}$. Furthermore,
the probabilities of the symmetric model are directly related to the
non-interacting system average populations by: \beqar
\label{eq:n_noninteracting}
n_{\uparrow}\big\vert_{\substack{U=0}}
&=&n_{\downarrow}\big\vert_{U=0}=2p_{\uparrow}\big\vert_{\substack{U=-2\varepsilon}}=2p_{\downarrow}\big\vert_{\substack{U=-2\varepsilon}}
= \frac{f(\varepsilon-\mu)+f(\varepsilon+\mu)}{2} \\ \nonumber
(1-n_{\uparrow})\big\vert_{\substack{U=0}} &=&
(1-n_{\downarrow})\big\vert_{U=0}=2p_{0}\big\vert_{\substack{U=-2\varepsilon}}=2p_{f}\big\vert_{\substack{U=-2\varepsilon}}
= 1-\frac{f(\varepsilon-\mu)+f(\varepsilon+\mu)}{2} .  \eeqar One
additional consequence of having the same electric currents is that
the steady-state entropy production will also coincide exactly for the
two models at this special symmetric point for $\varepsilon_f=0$.
\par The equivalence between the interacting and noninteracting
currents can also be derived within the EOM-NEGF approach. For
clarity, we restrict the discussion to the wide band limit and assume
that $\Gamma_{\rm L}=\Gamma_{\rm R}=\Gamma/2$. In this limit, we find
that the difference between the steady state currents of the
interacting and noninteracting models is given by (see supporting
information):
\begin{equation}
I_{\rm L}^{\left(U=-2\varepsilon\right)} - I_{\rm L}^{\left(U=0\right)} = e\Gamma^{2}U\frac{\left\langle n_{\uparrow}\right\rangle + \left\langle n_{\downarrow}\right\rangle}{2} \int_{-\infty}^{\infty}\frac{d\omega}{2\pi}\frac{2\omega\left(f^{\rm L}\left(\omega-\mu_{\rm L}\right)-f^{\rm R}\left(\omega-\mu_{\rm R}\right)\right)}{\omega^{4}+2\omega^{2}\left(\left(\frac{\Gamma}{2}\right)^{2}-\left(\frac{U}{2}\right)^{2}\right)+\left(\left(\frac{\Gamma}{2}\right)^{2}+\left(\frac{U}{2}\right)^{2}\right)^{2}}.
\end{equation}
When the electric bias is centred around $\varepsilon_{f}=0$ with $T_{\rm L}=T_{\rm R}$, the above difference vanishes exactly, as the integrand becomes an odd function of $\omega$. 

\section{Conclusions}
\label{sec:conclusions}
In conclusion, we have shown that the current in the Coulomb
  blockade regime for a symmetric Anderson impurity model is identical
  to the current in a non-interacting model. This was demonstrated
(numerically and analytically) within the quantum master equation
approach as well as the equation of motion NEGF approach (within a
two-particle closure) for a wide range of model parameters, bias
voltages, and temperatures.  The Anderson impurity model is the
canonical model to study weakly and strongly correlated effects away
from equilibrium and is routinely used to assess the accuracy of
approximate methods to compute the dynamics and steady-state
properties. Limiting such assessments to the special symmetric point,
as described in this work, is by no means a signature of a valid
many-body approximation. Such methods should always be tested away
from the symmetric case.

\begin{acknowledgement}
The authors would like to thank Wenjie Dou and Guy Cohen for fruitful
discussions. M.T. thanks Eran Rabani for kindly
hosting his sabbatical stay at the Chemistry Department of the
University of California at Berkeley and acknowledges support
by the German Research Foundation (DFG). J.O. acknowledges support from the Georg H. Endress foundation. This work was supported by the U.S. Department of Energy,
Office of Science, Office of Basic Energy Sciences, Materials Sciences
and Engineering Division, under Contract No. DEAC02-05-CH11231 within
the Physical Chemistry of Inorganic Nanostructures Program (KC3103).

\end{acknowledgement}

\begin{suppinfo}

In this supporting information we give a detailed derivation of the master equation and the equation of motion approach to the nonequilibrium Green's functions for the Anderson impurity model and assess the validity of these approximations.
\par
The Anderson impurity model is defined by the Hamiltonian
$H=H_{S}+H_{B}+V$, where
\beqar
H_{S}= & \sum_{\sigma}\varepsilon_{\sigma}d_{\sigma}^{\dagger}d_{\sigma}+Ud_{\uparrow}^{\dagger}d_{\uparrow}d_{\downarrow}^{\dagger}d_{\downarrow}
\eeqar
describes the impurity (or dot), referred to simply as the `system Hamiltonian',
\beqar
H_{B}= & \sum_{\underset{k\in L,R}{\sigma=\uparrow,\downarrow}}\varepsilon_{k}c_{k\sigma}^{\dagger}c_{k\sigma}
\eeqar
describes the noninteracting fermionic baths (or leads), and 
\beqar
V= & \sum_{\underset{k\in L,R}{\sigma=\uparrow,\downarrow}}t_{k}d_{\sigma}^{\dagger}c_{k\sigma}+{\rm h.c.},
\eeqar
describes the hybridization between the system and the leads. In the above, $d_{\sigma}^{\dagger}\ \left(d_{\sigma}\right)$ are the creation (annihilation) operators of an electron on the dot with spin $\sigma=\uparrow,\downarrow$ with one-body energy of $\varepsilon_{\sigma}$, $U$ is the on-site Hubbard interaction, $c_{k\sigma}^{\dagger}\ \left(c_{k\sigma}\right)$ are the creation (annihilation) operators of an electron in mode $k$ of the leads with energy $\varepsilon_{k}$, and $t_{k}$ is the hybridization between the dot and mode $k$ in the lead. The coupling to the quasi-continuous leads is modeled by
the leads' spectral function, $\Gamma_{\ell}\left(\omega\right)$, where
$t_{k\in\ell}=\sqrt{\Gamma_{\ell}\left(\ensuremath{\varepsilon}_{k}\right)\Delta\omega/2\pi}$ is the coupling between the dot and the $k$th mode of the $\ell=L,R$ bath, and $\Delta\omega$ is the discretization of the leads energy spectrum. Throughout, we take Planck's constant and the Boltzmann factor to be 1.  

\section{Master equation approach}

For weak coupling to the leads and at high temperatures, the master equation (ME) approach is an adequate approximation to describe transport in the Anderson impurity model.
In the many-body state basis of the system, the ME reduces to the rate equation: 
\beq
\label{eq:ME1}
\left(\begin{array}{c}
\dot{p}_{0}\\
\dot{p}_{\uparrow}\\
\dot{p}_{\downarrow}\\
\dot{p}_{f}
\end{array}\right)=M\left(\begin{array}{c}
p_{0}\\
p_{\uparrow}\\
p_{\downarrow}\\
p_{f}
\end{array}\right),
\eeq
where $p_0$ is the probability of occupying the empty dot state $\ket{0}$ , $p_{\uparrow}$ and $p_{\downarrow}$ are the probabilities of occupying a single electron in the states $\ket{\uparrow}$ and $\ket{\downarrow}$ respectively, and $p_f$ is the probability of having a fully occupied dot corresponding to the state $\ket{f}$.     
The transition matrix is given by :
\beq
\label{eq:ME2}
M = \sum_{\ell\in L,R}\left(\begin{array}{cccc}
-\left(\gamma^{\ell}_{\uparrow0}+\gamma^{\ell}_{\downarrow0}\right) & \gamma^{\ell}_{0\uparrow} & \gamma^{\ell}_{0\downarrow} & 0\\
\gamma^{\ell}_{\uparrow0} & -\left(\gamma^{\ell}_{0\uparrow}+\gamma^{\ell}_{f\uparrow}\right) & 0 & \gamma^{\ell}_{\uparrow f}\\
\gamma^{\ell}_{\downarrow0} & 0 & -\left(\gamma^{\ell}_{0\downarrow}+\gamma^{\ell}_{f\downarrow}\right) & \gamma_{\downarrow f}\\
0 & \gamma^{\ell}_{f\uparrow}& \gamma^{\ell}_{f\downarrow} & -\left(\gamma^{\ell}_{\uparrow f}+\gamma^{\ell}_{\downarrow f}\right)
\end{array}\right).
\eeq
Here $\gamma^{\ell}_{ij}$ is the transition rate from state $j$ to state $i$ induced by the $\ell=L,R$ bath, which can be calculated using Fermi's golden rule. 
Assuming symmetry between spin up and spin down and in the wide-band approximation, the rates are given explicitly by:
\beqar
\label{eq:ME3}
\gamma^{\ell}_{\uparrow0}&=&\gamma^{\ell}_{\downarrow0} \equiv \gamma^{\ell}(\varepsilon-\mu_{\ell}) =\Gamma_{\ell} f^{\ell}(\varepsilon-\mu_{\ell}) \\ \nonumber
\gamma^{\ell}_{0\uparrow}&=&\gamma^{\ell}_{0\downarrow} \equiv \gamma^{\ell}(-\varepsilon+\mu_{\ell}) =\Gamma_{\ell} f^{\ell}(-\varepsilon+\mu_{\ell}) \\ \nonumber
\gamma^{\ell}_{f\uparrow}&=&\gamma^{\ell}_{f\downarrow}\equiv \gamma^{\ell}(\varepsilon+U-\mu_{\ell})=\Gamma_{\ell} f^{\ell}(\varepsilon+U-\mu_{\ell})\\ \nonumber
\gamma^{\ell}_{\uparrow f}&=&\gamma^{\ell}_{\downarrow f}\equiv \gamma^{\ell}(-\varepsilon-U+\mu_{\ell})=\Gamma_{\ell} f^{\ell}(-\varepsilon-U+\mu_{\ell}),
\eeqar 
where $\mu_{\ell}$ is the chemical potential for the left ($\ell=L$) or right ($\ell=R$) leads.
\par
The electric, the energy and the heat currents from the $\ell$-bath are given by
\beqar
\label{eq:currentSI}
I_{\ell}(t)&=&\frac{e}{2}\sum_{ij}\left(N_{i}-N_{j}\right)j_{ij}^{\ell} \\ \nonumber
I^{E}_{\ell}(t)&=&\frac{1}{2}\sum_{ij}\left(E_{i}-E_{j}\right)j_{ij}^{\ell}\\ \nonumber
I^{H}_{\ell}(t)&=&I^{E}_{\ell}(t)-\mu_{\ell}I_{\ell}(t)
.
\eeqar
where $\quad j_{ij}^{\ell}=\gamma_{ij}^{\ell}p_{j}(t)-\gamma_{ji}^{\ell}p_{i}(t)$.
Here $N_{i}$ and $E_{i}$ are the number of electrons and the energy of the many-body state $i$, and $j_{ij}^{\ell}$ is the probability current from state $j$ to state $i$ induced by the coupling to the $\ell$ bath.
The probability currents can then be expressed as:
\beqar
\label{eq:pro_current}
j_{\uparrow0}^{\ell}&=&\gamma_{\uparrow0}^{\ell}(\varepsilon-\mu_{\ell})p_{0}(t)-\gamma_{\uparrow0}^{\ell}(-\varepsilon+\mu_{\ell})p_{\uparrow}(t)\\ \nonumber
j_{\downarrow0}^{\ell}&=&\gamma_{\downarrow0}^{\ell}(\varepsilon-\mu_{\ell})p_{0}(t)-\gamma_{\downarrow0}^{\ell}(-\varepsilon+\mu_{\ell})p_{\downarrow}(t)\\ \nonumber
j_{f\uparrow}^{\ell}&=&\gamma_{f\uparrow}^{\ell}(\varepsilon+U-\mu_{\ell})p_{\uparrow}(t)-\gamma_{f\uparrow}^{\ell}(-\varepsilon-U+\mu_{\ell})p_{f}(t)\\ \nonumber
j_{f\downarrow}^{\ell}&=&\gamma_{f\downarrow}^{\ell}(\varepsilon+U-\mu_{\ell})p_{\downarrow}(t)-\gamma_{f\downarrow}^{\ell}(-\varepsilon-U+\mu_{\ell})p_{f}(t).
\eeqar
\par
Note that if the left and right temperature of the baths are equal $(T_L=T_R)$ and the electric currents of the two systems  are the same, then  the entropy production $\Delta S$ of the two systems at steady-state is also identical. 
At steady-state $\Delta S= - \sum_{\ell}\frac{I_{\ell}^H}{T_\ell}$.  
Since $T_L=T_R$ the entropy production depends solely on the electric current.
\par
In the following we assume $\mu_L=-\mu_R\equiv\mu$.
\subsection{Noninteracting model, $U=0$:}  
Solving equations (\ref{eq:ME1})-(\ref{eq:ME3}) for the noninteracting system, the steady-state probabilities are given by:
\beqar
p_0(\infty)&=&\frac{\left(\gamma^L(-\varepsilon+\mu)+\gamma^R(-\varepsilon-\mu)\right)^2}{\left(\Gamma_L+\Gamma_R\right)^2} \\ \nonumber
p_f(\infty)&=&\frac{\left(\gamma^L(\varepsilon-\mu)+\gamma^R(\varepsilon+\mu)\right)^2}{\left(\Gamma_L+\Gamma_R\right)^2} \\ \nonumber
p_{\uparrow /\downarrow}(\infty)&=&\frac{\left(\gamma^L(-\varepsilon+\mu)+\gamma^R(-\varepsilon-\mu)\right)\left(\gamma^L(\varepsilon-\mu)+\gamma^R(\varepsilon+\mu)\right)}{\left(\Gamma_L+\Gamma_R\right)^2}.
\eeqar
Since spin up and down are independent, the single particle probabilities are simply $q_0=\sqrt{p_0}$ of having no electron and $q_1=\sqrt{p_f}$ of having a single electron. Thus, the occupation number (average population) is given by,
\beq
\label{eq:n_noninteractingSI}
n_{\uparrow} =n_{\downarrow}=\frac{\gamma^L(\varepsilon-\mu)+\gamma^R(\varepsilon+\mu)}{\Gamma^L+\Gamma^R}\bigg\vert_{\substack{T_l=T_r \\ \Gamma_L=\Gamma_R}} = \frac{f(\varepsilon-\mu)+f(\varepsilon+\mu)}{2} .
\eeq   
The steady-state current can be calculated using equations (\ref{eq:currentSI}) and (\ref{eq:pro_current}), and is given by
\beq
\label{eq:noninteracting_currentSI}
I_L(\infty)=-I_R(\infty)=\frac{2\Gamma_L \Gamma_R }{\Gamma_L + \Gamma_R}\left(f^L(\varepsilon-\mu)-f^R(\varepsilon+\mu) \right).
\eeq 
\subsection{Interacting model, symmetric Anderson model $U=-2\varepsilon$:}  
The steady state probabilities from equations (\ref{eq:ME1})-(\ref{eq:ME3}) are
\beqar
\label{eq:pop_symmetric}
p_0(\infty)&=&\frac{\left(\gamma^L(\varepsilon+\mu)+\gamma^R(\varepsilon-\mu)\right)\left(\Gamma_L+\Gamma_R -\gamma^L(\varepsilon-\mu)-\gamma^R(\varepsilon+\mu)\right)}{\left(\Gamma_L+\Gamma_R\right)\left(\gamma^L(\varepsilon-\mu)+\gamma^L(\varepsilon+\mu) +\gamma^R(\varepsilon-\mu)+\gamma^R(\varepsilon+\mu)\right) } \\ \nonumber
p_f(\infty)&=&\frac{\left(\Gamma_L+\Gamma_R - \gamma^L(\varepsilon+\mu)-\gamma^R(\varepsilon - \mu)\right)\left(\gamma^L(\varepsilon-\mu)+\gamma^R(\varepsilon+\mu)\right)}{\left(\Gamma_L+\Gamma_R\right)\left(\gamma^L(\varepsilon-\mu)+\gamma^L(\varepsilon+\mu) +\gamma^R(\varepsilon-\mu)+\gamma^R(\varepsilon+\mu)\right) } \\ \nonumber
p_{\uparrow /\downarrow}(\infty)&=&\frac{\left(\gamma^L(\varepsilon+\mu)+\gamma^R(\varepsilon-\mu)\right)\left(\gamma^L(\varepsilon-\mu)+\gamma^R(\varepsilon+\mu)\right)}{\left(\Gamma_L+\Gamma_R\right)\left(\gamma^L(\varepsilon-\mu)+\gamma^L(\varepsilon+\mu) +\gamma^R(\varepsilon-\mu)+\gamma^R(\varepsilon+\mu)\right)}.
\eeqar
The additional necessary condition for having half-filling is that $\Gamma_L=\Gamma_R$ and $T_L=T_R$. In this case, Eq. (\ref{eq:pop_symmetric}) can be simplified to yield:
\beqar
p_0(\infty)&=& p_f(\infty)=\frac{f(-\varepsilon+\mu)+f(-\varepsilon-\mu)}{4} \\ \nonumber
p_{\uparrow}(\infty)&=&  p_{\downarrow}(\infty) =\frac{f(\varepsilon-\mu)+f(\varepsilon+\mu)}{4},
\eeqar
and the occupation number is $n_{\uparrow}=n_{\downarrow}=\frac{1}{2}$. 
The probabilities $q_0$ and $q_1$ of the noninteracting system  are exactly twice the probabilities of the symmetric interacting system, i.e., $q_0=2p_0=2p_f$ and  $q_1=2p_{\uparrow}=2p_{\downarrow}$.
\par
Using equations (\ref{eq:currentSI}), (\ref{eq:pro_current}) and (\ref{eq:pop_symmetric}) the steady-state current from the left lead is,
\beq
I_L(\infty)=\frac{2\Gamma_L\Gamma_R\left(f^L(\varepsilon-\mu)f^R(\varepsilon-\mu)-f^L(\varepsilon+\mu)f^R(\varepsilon+\mu) \right)}{\Gamma_L \left( f^L(\varepsilon-\mu )+f^L(\varepsilon+\mu)\right)+\Gamma_R \left( f^R(\varepsilon-\mu )+f^R(\varepsilon+\mu)\right)}
\eeq 
For $T_L=T_R\equiv T $ the steady-state current takes the form
\beq
I_L(\infty)=-I_R(\infty)=\frac{2\Gamma_L \Gamma_R }{\Gamma_L + \Gamma_R}\left(f(\varepsilon-\mu)-f(\varepsilon+\mu) \right),
\eeq  
which is identical to the current in the noninteracting model Eq. (\ref{eq:noninteracting_currentSI}). 
\par
In summary, the conditions for having equal currents for the interacting and noninteracting systems are: equal temperatures of the left and right leads $T_L=T_R$, Fermi energy equal to zero $\mu_L=-\mu_R$ (symmetric distribution of the bias) and $U=-2\varepsilon$. 
Note that the leads' spectral functions, $\Gamma_L$ and $\Gamma_R$, are not necessarily equal as required for obtaining half-filling, but must be an even function in energy.

As a side note we mention that the difference between the currents in the symmetric and noninteracting models may be expanded to first order in the temperature difference between the leads, $\Delta T=T_R-T_L$, to obtain
\beq
I_L^{\left(U=-2\varepsilon\right)} - I_L^{\left(U=0\right)}=\frac{\Gamma\left( \varepsilon+\varepsilon \cosh\left(\frac{\varepsilon}{T_L}\right) \cosh\left(\frac{\mu}{T_L}\right) -\mu\sinh\left(\frac{\varepsilon}{T_L}\right) \sinh\left(\frac{\mu}{T_L}\right) \right)}{2\left( \cosh\left(\frac{\varepsilon}{T_L}\right)+ \cosh\left(\frac{\mu}{T_L}\right)\right)^2} \frac{\Delta T}{T_L^2}.
\eeq
Thus, one may use the noninteracting result to obtain the interacting current in the vicinity of the symmetry point.

\section{Nonequilibrium Green's function approach}

We begin by defining the impurity one-body nonequilibrium Green's
function (GF) on the Keldysh contour, \cite{Schwinger1961,Keldysh1964}
\begin{equation}
G_{\sigma}\left(\tau,\tau'\right)=-i\left\langle \hat{T}_{c}\left[d_{\sigma}\left(\tau\right)d_{\sigma}^{\dagger}\left(\tau'\right)\right]\right\rangle 
\end{equation}
where $\hat{T}_{c}$ is the contour ordering operator. Using the equation of motion (EOM) technique \cite{Zubarev1960,Bruevich1962,Scheck2013}, we take the derivative of the GF according to one of the contour times, producing the following equation
\begin{equation}
i\frac{\partial G_{\sigma}\left(\tau,\tau'\right)}{\partial\tau}=\delta_{c}\left(\tau-\tau'\right)+\varepsilon_{\sigma}G_{\sigma}\left(\tau,\tau'\right)+UG_{\sigma}^{\left(2\right)}\left(\tau,\tau'\right)+\sum_{k\in L,R}t_{k}F_{k\sigma}\left(\tau,\tau'\right),
\end{equation}
where we defined
\begin{align}
G_{\sigma}^{\left(2\right)}\left(\tau,\tau'\right) & =-i\left\langle \hat{T}_{c}\left[d_{\sigma}\left(\tau\right)n_{\bar{\sigma}}\left(\tau\right)d_{\sigma}^{\dagger}\left(\tau'\right)\right]\right\rangle ,\\
F_{k\sigma}\left(\tau,\tau'\right) & =-i\left\langle \hat{T}_{c}\left[c_{k\sigma}\left(\tau\right)d_{\sigma}^{\dagger}\left(\tau'\right)\right]\right\rangle .
\end{align}
The above satisfy the following equations of motion
\begin{align}
i\frac{\partial G_{\sigma}^{\left(2\right)}\left(\tau,\tau'\right)}{\partial\tau'} & =-\delta\left(\tau-\tau'\right)\left\langle n_{\bar{\sigma}}\left(\tau\right)\right\rangle -\varepsilon_{\sigma}G_{\sigma}^{\left(2\right)}\left(\tau,\tau'\right)-UG_{\sigma}^{\left(3\right)}\left(\tau,\tau'\right)-\sum_{k\in L,R}t_{k}^{*}F_{k\sigma}^{\left(2\right)}\left(\tau,\tau'\right),\\ 
i\frac{\partial F_{k\sigma}}{\partial\tau}\left(\tau,\tau'\right) & =\varepsilon_{k\sigma}F_{k\sigma}\left(\tau,\tau'\right)+t_{k}^{*}G_{\sigma}\left(\tau,\tau'\right),
\end{align}
where we also defined 
\begin{align}
G_{\sigma}^{\left(3\right)}\left(\tau,\tau'\right) & =-i\left\langle \hat{T}_{c}\left[d_{\sigma}\left(\tau\right)n_{\bar{\sigma}}\left(\tau\right)d_{\sigma}^{\dagger}\left(\tau'\right)n_{\bar{\sigma}}\left(\tau'\right)\right]\right\rangle ,\\
F_{k\sigma}^{\left(2\right)}\left(\tau,\tau'\right) & =-i\left\langle \hat{T}_{c}\left[d_{\sigma}\left(\tau\right)n_{\bar{\sigma}}\left(\tau\right)c_{k\sigma}^{\dagger}\left(\tau'\right)\right]\right\rangle .
\end{align}
Since the Anderson model is not analytically solvable, this ever-growing
set of inter-dependent equations will not come to a close, and so,
we choose an appropriate closure, which is known to be a good approximation in the Coulomb blockade regime.
Within this closure, the equation of motion for $F_{k\sigma}^{\left(2\right)}\left(\tau,\tau'\right)$
is taken exactly,
\begin{align}
i\frac{\partial F_{k\sigma}^{\left(2\right)}\left(\tau,\tau'\right)}{\partial\tau'} & =-\varepsilon_{k\sigma}F_{k\sigma}^{\left(2\right)}\left(\tau,\tau'\right)-t_{k}G_{\sigma}^{\left(2\right)}\left(\tau,\tau'\right),
\end{align}
whereas $G_{\sigma}^{\left(3\right)}\left(\tau,\tau'\right)$ is treated
approximately. If one derives the EOM for $G_{\sigma}^{\left(2\right)}\left(\tau,\tau'\right)$
under the noninteracting, uncoupled Hamiltonian, the resulting equation of motion is found to be identical to that of $G_{\sigma}^{\left(3\right)}\left(\tau,\tau'\right)$ under the same Hamiltonian (and has the same initial value).
Therefore, within this closure, we will use the approximation:
\begin{equation}
G_{\sigma}^{\left(3\right)}\left(\tau,\tau'\right)\simeq G_{\sigma}^{\left(2\right)}\left(\tau,\tau'\right).
\end{equation}
In this derivation, we will also consider the neglected terms from
the closure, in order to keep the full description in mind. Thus,
we define
\begin{equation}
G_{\sigma}^{\left(3\right)}=G_{\sigma}^{\left(2\right)}+\Gamma F_{\sigma}^{\left({\rm corr}\right)},
\end{equation}
where $\Gamma F_{\sigma}^{\left({\rm corr}\right)}$ is the difference between the exact $G_{\sigma}^{\left(3\right)}$ and our approximation and the leading contribution due to the coupling to the leads is factored out. As a side note, we can show that the correction term will contribute at least $\propto\Gamma$ by writing out the full equation of motion for $G_\sigma^{\left(3\right)}$,
\begin{eqnarray*}
i\frac{\partial G_{\sigma}^{\left(3\right)}\left(\tau,\tau'\right)}{\partial\tau} & = & \frac{\partial}{\partial\tau}\left[\theta\left(\tau-\tau'\right)\left\langle \left\{ d_{\sigma}\left(\tau\right)n_{\bar{\sigma}}\left(\tau\right),d_{\sigma}^{\dagger}\left(\tau'\right)n_{\bar{\sigma}}\left(\tau'\right)\right\} \right\rangle \right]\\
 & = & \delta\left(\tau-\tau'\right)\left\langle n_{\bar{\sigma}}\left(\tau\right)\right\rangle +\left\langle \hat{T}_{c}\left[\dot{d}_{\sigma}\left(\tau\right)n_{\bar{\sigma}}\left(\tau\right)d_{\sigma}^{\dagger}\left(\tau'\right)n_{\bar{\sigma}}\left(\tau'\right)\right]\right\rangle\\ & & +\left\langle \hat{T}_{c}\left[d_{\sigma}\left(\tau\right)\dot{n}_{\bar{\sigma}}\left(\tau\right)d_{\sigma}^{\dagger}\left(\tau'\right)n_{\bar{\sigma}}\left(\tau'\right)\right]\right\rangle \\
 & = & \delta\left(\tau-\tau'\right)\left\langle n_{\bar{\sigma}}\left(\tau\right)\right\rangle \\
 &  & -i\left\langle \hat{T}_{c}\left[\left(\varepsilon_{\sigma}d_{\sigma}\left(\tau\right)+Ud_{\sigma}\left(\tau\right)n_{\bar{\sigma}}\left(\tau\right)+\sum_{k\in L,R}t_{k}c_{k\sigma}\left(\tau\right)\right)n_{\bar{\sigma}}\left(\tau\right)d_{\sigma}^{\dagger}\left(\tau'\right)n_{\bar{\sigma}}\left(\tau'\right)\right]\right\rangle \\
 &  & -i\left\langle \hat{T}_{c}\left[d_{\sigma}\left(t\right)\left(\sum_{k\in L,R}t_{k}d_{\sigma}^{\dagger}\left(t\right)c_{k\sigma}\left(t\right)+\sum_{k\in L,R}t_{k}d_{\sigma}\left(t\right)c_{k\sigma}^{\dagger}\left(t\right)\right)d_{\sigma}^{\dagger}\left(t'\right)n_{\bar{\sigma}}\left(t'\right)\right]\right\rangle .
 \end{eqnarray*}
To arrive at the uncoupled equation, we neglect all terms with $\sum_{k\in L,R}t_{k}$, keeping only terms that appear in the propagation under the dot Hamiltonian,
 \begin{eqnarray*}
i\frac{\partial G_{\sigma}^{\left(3\right)}\left(\tau,\tau'\right)}{\partial\tau} & \simeq & \delta\left(\tau-\tau'\right)\left\langle n_{\bar{\sigma}}\left(\tau\right)\right\rangle -i\varepsilon_{\sigma}\left\langle \hat{T}_{c}\left[d_{\sigma}\left(\tau\right)n_{\bar{\sigma}}\left(\tau\right)d_{\sigma}^{\dagger}\left(\tau'\right)n_{\bar{\sigma}}\left(\tau'\right)\right]\right\rangle \\ & &-iU\left\langle \hat{T}_{c}\left[d_{\sigma}\left(\tau\right)n_{\bar{\sigma}}\left(\tau\right)n_{\bar{\sigma}}\left(\tau\right)d_{\sigma}^{\dagger}\left(\tau'\right)n_{\bar{\sigma}}\left(\tau'\right)\right]\right\rangle \\
 & = & \delta\left(\tau-\tau'\right)\left\langle n_{\bar{\sigma}}\left(\tau\right)\right\rangle +\left(\varepsilon_{\sigma}+U\right)G_{\sigma}^{\left(3\right)}\left(\tau,\tau'\right)
\end{eqnarray*}
thus,
\begin{align}
G_{\sigma}^{\left(3\right)}\left(\tau,\tau'\right) & =G_{\sigma}^{\left(2\right)}\left(\tau,\tau'\right)+\int_{0}^{\tau}\left[{\rm neglected\ terms}\right]\left(\tau_{1},\tau'\right){\rm d}\tau_{1}\\
 & \equiv G_{\sigma}^{\left(2\right)}\left(\tau,\tau'\right)+\Gamma F_{\sigma}^{\left({\rm corr}\right)}.\nonumber 
\end{align}

Solving the equations of motion on the Keldysh contour, one arrives
at the following integral equations
\begin{align}
G_{\sigma}\left(\tau,\tau'\right)= & \,g_{\sigma}^{(0)}\left(\tau,\tau'\right)+\iint{\rm d}\tau_{1}{\rm d}\tau_{2}\,g_{\sigma}^{(0)}\left(\tau,\tau_{1}\right)\Sigma_{\ell\sigma}\left(\tau_{1},\tau_{2}\right)G_{\sigma}\left(\tau_{2},\tau'\right)\\ \nonumber
 & +U\int{\rm d}\tau_{1}\,g_{\sigma}^{(0)}\left(\tau,\tau_{1}\right)G_{\sigma}^{\left(2\right)}\left(\tau_{1},\tau'\right),\\
G_{\sigma}^{\left(2\right)}\left(\tau,\tau'\right)= & \,g_{\sigma}^{\left(U\right)}\left(\tau,\tau'\right)\left\langle n_{\bar{\sigma}}\left(\tau\right)\right\rangle +\iint{\rm d}\tau_{1}{\rm d}\tau_{2}\,G_{\sigma}^{\left(2\right)}\left(\tau,\tau_{2}\right)\Sigma_{\ell\sigma}\left(\tau_{2},\tau_{1}\right)g_{\sigma}^{\left(U\right)}\left(\tau_{1},\tau'\right)\\
 & +U\Gamma\int{\rm d}\tau_{1}\,g_{\sigma}^{\left(U\right)}\left(\tau,\tau_{1}\right)F_{\sigma}^{\left({\rm corr}\right)}\left(\tau_{1},\tau'\right),\nonumber 
\end{align}
where the noninteracting Green's function on the contour are defined as
\begin{align}
g_{\sigma}^{(0)}\left(\tau,\tau'\right) & =-i\left\langle \hat{T}_{c}\left[d_{\sigma}\left(\tau\right)d_{\sigma}^{\dagger}\left(\tau'\right)\right]\right\rangle _{{\rm under}\ H_{0}=\varepsilon_{\sigma}n_{\sigma}},\\ \nonumber
g_{\sigma}^{\left(U\right)}\left(\tau,\tau'\right) & =-i\left\langle \hat{T}_{c}\left[d_{\sigma}\left(\tau\right)d_{\sigma}^{\dagger}\left(\tau'\right)\right]\right\rangle _{{\rm under}\ H_{0}=\left(\varepsilon_{\sigma}+U\right)n_{\sigma}},
\end{align}
and the self energy due to the coupling to the leads (identical to the
noninteracting model self energy) is given by
\begin{equation}
\Sigma_{\ell\sigma}\left(\tau,\tau'\right)=-i\sum_{k\in L,R}\left|t_{k}\right|^{2}\left\langle \hat{T}_{c}\left[c_{k\sigma}\left(\tau\right)c_{k\sigma}^{\dagger}\left(\tau'\right)\right]\right\rangle _{{\rm under}\ H_{0}=H_{B}}.
\end{equation}

By means of analytical continuation (Langreth rules)\cite{Langreth76}, we find that the equations for the retarded and lesser Green's functions are given by
\begin{align}
G_{\sigma}^{r}\left(t,t'\right)= & \,g_{\sigma}^{(0)r}\left(t-t'\right)+\iint{\rm d}t_{1}{\rm d}t_{2}\,g_{\sigma}^{(0)r}\left(t-t_{1}\right)\Sigma_{\ell\sigma}^{r}\left(t_{1}-t_{2}\right)G_{\sigma}^{r}\left(t_{2},t'\right)\\
 & +U\int{\rm d}t_{1}\,g_{\sigma}^{(0)r}\left(t-t_{1}\right)G_{\sigma}^{\left(2\right)r}\left(t_{1},t'\right),\nonumber \\
G_{\sigma}^{\left(2\right)r}\left(t,t'\right)= & \,g_{\sigma}^{\left(U\right)r}\left(t-t'\right)\left\langle n_{\bar{\sigma}}\left(t\right)\right\rangle +\iint{\rm d}t_{1}{\rm d}t_{2}\,G_{\sigma}^{\left(2\right)r}\left(t,t_{2}\right)\Sigma_{\ell\sigma}^{r}\left(t_{2}-t_{1}\right)g_{\sigma}^{\left(U\right)r}\left(t_{1}-t'\right)\\
 & +U\Gamma\int g_{\sigma}^{\left(U\right)r}\left(t-t_{1}\right)F_{\sigma}^{\left({\rm corr}\right)r}\left(t_{1},t'\right)\nonumber \\
G_{\sigma}^{<}\left(t,t'\right)= & g_{\sigma}^{(0)<}\left(t-t'\right)+\iint g_{\sigma}^{(0)r}\left(t-t_{1}\right)\Sigma_{\ell\sigma}^{r}\left(t_{1}-t_{2}\right)G_{\sigma}^{<}\left(t_{2},t'\right)
\\ \nonumber
& +\iint g_{\sigma}^{(0)r}\left(t-t_{1}\right)\Sigma_{\ell\sigma}^{<}\left(t_{1}-t_{2}\right)G_{\sigma}^{a}\left(t_{2},t'\right) +\iint g_{\sigma}^{(0)<}\left(t-t_{1}\right)\Sigma_{\ell\sigma}^{a}\left(t_{1}-t_{2}\right)G_{\sigma}^{a}\left(t_{2},t'\right)\\
 & +U\int g_{\sigma}^{(0)r}\left(t-t_{1}\right)G_{\sigma}^{\left(2\right)<}\left(t_{1},t'\right)\nonumber  +U\int g_{\sigma}^{(0)<}\left(t-t_{1}\right)G_{\sigma}^{\left(2\right)a}\left(t_{1},t'\right),\nonumber \\
G_{\sigma}^{\left(2\right)<}\left(t,t'\right)= & g_{\sigma}^{\left(U\right)<}\left(t-t'\right)\left\langle n_{\bar{\sigma}}\left(t\right)\right\rangle +\iint g_{\sigma}^{\left(U\right)r}\left(t-t_{1}\right)\Sigma_{\ell\sigma}^{r}\left(t_{1}-t_{2}\right)G_{\sigma}^{\left(2\right)<}\left(t_{2},t'\right)\\ \nonumber
 & +\iint g_{\sigma}^{\left(U\right)r}\left(t-t_{1}\right)\Sigma_{\ell\sigma}^{<}\left(t_{1}-t_{2}\right)G_{\sigma}^{\left(2\right)a}\left(t_{2},t'\right) \\ \nonumber
 & +\iint g_{\sigma}^{\left(U\right)<}\left(t-t_{1}\right)\Sigma_{\ell\sigma}^{a}\left(t_{1}-t_{2}\right)G_{\sigma}^{\left(2\right)a}\left(t_{2},t'\right)\nonumber \\
 & +U\Gamma\int g_{\sigma}^{\left(U\right)r}\left(t-t_{1}\right)F_{\sigma}^{\left({\rm corr}\right)<}\left(t_{1},t'\right)+U\Gamma\int g_{\sigma}^{\left(U\right)<}\left(t-t_{1}\right)F_{\sigma}^{\left({\rm corr}\right)a}\left(t_{1},t'\right),\nonumber 
\end{align}
where the noninteracting Green's functions are given by
\begin{align}
g_{\sigma}^{(0)r}\left(t\right) & =-i\theta\left(t\right)e^{-i\varepsilon_{\sigma}t},\\ \nonumber
g_{\sigma}^{(0)<}\left(t\right) & =ie^{-i\varepsilon_{\sigma}t}n_{\sigma}\left(0\right),\\ \nonumber
g_{\sigma}^{\left(U\right)r}\left(t\right) & =-i\theta\left(t\right)e^{-i\left(\varepsilon_{\sigma}+U\right)t},\\ \nonumber
g_{\sigma}^{\left(U\right)<}\left(t\right) & =ie^{-i\left(\varepsilon_{\sigma}+U\right)t}n_{\sigma}\left(0\right),\\ \nonumber
\Sigma_{\ell\sigma}^{r}\left(t\right) & =-i\theta\left(t\right)\int_{-\infty}^{\infty}\frac{\mbox{d}\omega}{2\pi}\left(\Gamma_{L}\left(\omega\right)+\Gamma_{R}\left(\omega\right)\right)e^{-\frac{i}{\hbar}\omega t}=-i\theta\left(t\right)\Gamma\left(t\right),\\ \nonumber
\Sigma_{\ell\sigma}^{<}\left(t\right) & =i\int_{-\infty}^{\infty}\frac{\mbox{d}\omega}{2\pi}\left(\Gamma_{L}\left(\omega\right)f_{F}\left(\omega-\mu_{L}\right)+\Gamma_{R}\left(\omega\right)f_{F}\left(\omega-\mu_{R}\right)\right)e^{-\frac{i}{\hbar}\omega t}.
\end{align}
\par
In steady state, all functions depend on time differences rather than two time variables, and thus, the integral equation are simplified to algebraic equations in frequency space. Therefore,
\begin{align}
\label{eq:steadystateGF}
G_{\sigma}^{r}\left(\omega\right) & =g_{\sigma}^{(0)r}\left(\omega\right)+g_{\sigma}^{(0)r}\left(\omega\right)\Sigma_{\sigma}^{r}\left(\omega\right)G_{\sigma}^{r}\left(\omega\right)+Ug_{\sigma}^{(0)r}\left(\omega\right)G_{\sigma}^{\left(2\right)r}\left(\omega\right),\\ \nonumber
G_{\sigma}^{\left(2\right)r}\left(\omega\right) & =g_{\sigma}^{\left(U\right)r}\left(\omega\right)\left\langle n_{\bar{\sigma}}\right\rangle +g_{\sigma}^{\left(U\right)r}\left(\omega\right)\Sigma_{\sigma}^{r}\left(\omega\right)G_{\sigma}^{\left(2\right)r}\left(\omega\right)+U\Gamma g_{\sigma}^{\left(U\right)r}\left(\omega\right)F_{\sigma}^{\left({\rm corr}\right)r}\left(\omega\right),\\ \nonumber
G_{\sigma}^{<}\left(\omega\right) & =g_{\sigma}^{(0)<}\left(\omega\right)+g_{\sigma}^{(0)r}\left(\omega\right)\Sigma_{\sigma}^{r}\left(\omega\right)G_{\sigma}^{<}\left(\omega\right)+g_{\sigma}^{(0)r}\left(\omega\right)\Sigma_{\sigma}^{<}\left(\omega\right)G_{\sigma}^{a}\left(\omega\right)+g_{\sigma}^{(0)<}\left(\omega\right)\Sigma_{\sigma}^{a}\left(\omega\right)G_{\sigma}^{a}\left(\omega\right)\\ \nonumber
 & +Ug_{\sigma}^{(0)r}\left(\omega\right)G_{\sigma}^{\left(2\right)<}\left(\omega\right)+Ug_{\sigma}^{(0)<}\left(\omega\right)G_{\sigma}^{\left(2\right)a}\left(\omega\right),\nonumber \\ \nonumber
G_{\sigma}^{\left(2\right)<}\left(\omega\right) & =g_{\sigma}^{\left(U\right)<}\left(\omega\right)\left\langle n_{\bar{\sigma}}\right\rangle +g_{\sigma}^{\left(U\right)r}\left(\omega\right)\Sigma_{\sigma}^{r}\left(\omega\right)G_{\sigma}^{\left(2\right)<}\left(\omega\right)+g_{\sigma}^{\left(U\right)r}\left(\omega\right)\Sigma_{\sigma}^{<}\left(\omega\right)G_{\sigma}^{\left(2\right)a}\left(\omega\right)\\
 & +g_{\sigma}^{\left(U\right)<}\left(\omega\right)\Sigma_{\sigma}^{a}\left(\omega\right)G_{\sigma}^{\left(2\right)a}\left(\omega\right)+U\Gamma g_{\sigma}^{\left(U\right)r}\left(\omega\right)F_{\sigma}^{\left({\rm corr}\right)<}\left(\omega\right)+U\Gamma g_{\sigma}^{\left(U\right)<}\left(\omega\right)F_{\sigma}^{\left({\rm corr}\right)a}\left(\omega\right),\nonumber 
\end{align}
where the $\sigma$-spin electron steady state population, $\left\langle n_{\sigma}\right\rangle $
is given by:
\begin{equation}
\left\langle n_{\sigma}\right\rangle =-i\int_{-\infty}^{\infty}\frac{d\omega}{2\pi}G_{\sigma}^{<}\left(\omega\right),
\end{equation}
and the noninteracting Green's functions in frequency are given by
\begin{align}
g_{\sigma}^{(0)r}\left(\omega\right) & =\frac{1}{\omega-\varepsilon_{\sigma}+i\eta},\\ \nonumber
g_{\sigma}^{(0)<}\left(\omega\right) & =2\pi\left\langle d^{\dagger}d\right\rangle _{0}\delta\left(\omega-\varepsilon_{\sigma}\right),\\ \nonumber
g_{\sigma}^{\left(U\right)r}\left(\omega\right) & =\frac{1}{\omega-\varepsilon_{\sigma}-U+i\eta},\\ \nonumber
g_{\sigma}^{\left(U\right)<}\left(\omega\right) & =2\pi\left\langle d^{\dagger}d\right\rangle _{0}\delta\left(\omega-\varepsilon_{\sigma}-U\right),\\ \nonumber
\Sigma_{\sigma}^{r}\left(\omega\right) & =-\frac{i}{2}\Gamma\left(\omega\right), \\ \nonumber
\Sigma_{\sigma}^{<}\left(\omega\right) & =i\left(\Gamma_{L}\left(\omega\right)f\left(\omega-\mu_{L}\right)+\Gamma_{R}\left(\omega\right)f\left(\omega-\mu_{R}\right)\right).
\end{align}
In the above, we assumed that the spectral function characterizing the dot-lead coupling is symmetric, i.e. $\Gamma\left(\omega\right)=\Gamma\left(-\omega\right)$.
Note we also set the Fermi energy of the leads to $\varepsilon_{f}=0$,
and consider spectral functions that are symmetric around $\varepsilon_{f}$.
This will be critical for considering the symmetric point of the Anderson model. The above equations may now be solved analytically. 
\subsection{Symmetric case - steady state solution}
Let us now consider the more special case of the symmetric Anderson model, where:
\begin{align}
\varepsilon_{\uparrow} & =\varepsilon_{\downarrow}=-U/2,\\ \nonumber
\varepsilon_{\sigma}+ & U=U/2.
\end{align}
The noninteracting GFs in steady state take the form:
\begin{align}
g_{\sigma}^{(0)r}\left(\omega\right) & =\frac{1}{\omega+U/2+i\eta},\\ \nonumber
g_{\sigma}^{(0)<}\left(\omega\right) & =2\pi\left\langle d^{\dagger}d\right\rangle _{0}\delta\left(\omega+U/2\right),\\ \nonumber
g_{\sigma}^{\left(U\right)r}\left(\omega\right) & =\frac{1}{\omega-U/2+i\eta},\\ \nonumber
g_{\sigma}^{\left(U\right)<}\left(\omega\right) & =2\pi\left\langle d^{\dagger}d\right\rangle _{0}\delta\left(\omega-U/2\right).
\end{align}

Solving the steady state equations in frequency space, we start with the retarded GF. Rewriting the first line of equation \ref{eq:steadystateGF}, we find
\begin{align}
G_\sigma^{r} & =g_\sigma^{(0)r}\left(1-g_\sigma^{(0)r}\Sigma^{r}\right)^{-1}+g_\sigma^{(0)r}\left(1-g_\sigma^{(0)r}\Sigma^{r}\right)^{-1}UG_\sigma^{\left(2\right)r},\\
 & =G_\sigma^{(0)r}+G_\sigma^{(0)r}UG_\sigma^{\left(2\right)r},\nonumber 
\end{align}
where we recognized the retarded GF for the noninteracting model \cite{meir_landauer_1992} with $\varepsilon=-\frac{U}{2}$,
\begin{equation}
\label{eq:Gr}
G_\sigma^{(0)r}=\left(\left(g_\sigma^{(0)r}\right)^{-1}-\Sigma^{r}\right)^{-1}=\left(\omega+\frac{U}{2}-\Sigma^{r}\right)^{-1}.
\end{equation}
Similarly, from the second line of \ref{eq:steadystateGF}, we find:
\begin{align}
\label{eq:G2r}
G_\sigma^{\left(2\right)r}\left(\omega\right) & =g_\sigma^{\left(U\right)r}\left(1-g_\sigma^{\left(U\right)r}\Sigma^{r}\right)^{-1}\left\langle n_{\bar{\sigma}}\right\rangle +g_\sigma^{\left(U\right)r}\left(1-g_\sigma^{(U)r}\Sigma^{r}\right)^{-1}U\Gamma F_{\sigma}^{\left({\rm corr}\right)r}\\
 & =G_\sigma^{(0')r}\left\langle n_{\bar{\sigma}}\right\rangle +G_\sigma^{(0')r}U\Gamma F_{\sigma}^{\left({\rm corr}\right)r},\nonumber 
\end{align}
where
\begin{equation}
G_\sigma^{(0')r}=\left(\left(g_\sigma^{\left(U\right)r}\right)^{-1}-\Sigma^{r}\right)^{-1}=\left(\omega-\frac{U}{2}-\Sigma^{r}\right)^{-1}
\end{equation}
is the retarded GF for the noninteracting model with $\varepsilon=+\frac{U}{2}$.

Plugging the above into equation (\ref{eq:Gr}), we find:
\begin{equation}
G_\sigma^{r}=G_\sigma^{(0)r}+G_\sigma^{(0)r}UG_\sigma^{(0')r}\left\langle n_{\bar{\sigma}}\right\rangle +G_\sigma^{\left({\rm corr}\right)r},
\end{equation}
where we defined the closure-correction contribution to $G^{r}$:
\begin{equation}
G_\sigma^{\left({\rm corr}\right)r}=U^{2}\Gamma G_\sigma^{(0)r}G_\sigma^{(0')r}F_\sigma^{\left({\rm corr}\right)r}.
\end{equation}
Finally, using
\begin{align}
G_\sigma^{(0)r}\cdot G_\sigma^{(0')r}= & \left(\omega+\frac{U}{2}-\Sigma^{r}\right)^{-1}\cdot\left(\omega-\frac{U}{2}-\Sigma^{r}\right)^{-1}\\
= & \frac{1}{\omega^{2}-\left(\frac{U}{2}\right)^{2}-2\omega\Sigma^{r}+\left(\Sigma^{r}\right)^{2}}\nonumber \\
= & \frac{1}{\left(\omega-i\frac{\Gamma}{2}\right)^{2}-\left(\frac{U}{2}\right)^{2}},\nonumber 
\end{align}
we find
\begin{align}
G_{\sigma}^{r}-G_{\sigma}^{(0)r} & =G_\sigma^{(0)r}UG_\sigma^{(0')r}\left\langle n_{\bar{\sigma}}\right\rangle +G_{\sigma}^{\left({\rm corr}\right)r}\\ \nonumber
 & =\frac{U\left\langle n_{\bar{\sigma}}\right\rangle }{\left(\omega-i\frac{\Gamma}{2}\right)^{2}-\left(\frac{U}{2}\right)^{2}}+G_{\sigma}^{\left({\rm corr}\right)r}
\end{align}
where
\begin{equation}
G_{\sigma}^{\left({\rm corr}\right)r}=G_\sigma^{(0)r}UG_\sigma^{(0')r}U\Gamma F_{\sigma}^{\left({\rm corr}\right)r}=\frac{U^{2}\Gamma F_{\sigma}^{\left({\rm corr}\right)r}}{\left(\omega-i\frac{\Gamma}{2}\right)^{2}-\left(\frac{U}{2}\right)^{2}},
\end{equation}
and we took the wide-band limit (WBL) for the lead-dot coupling, $\Sigma^{r}\left(\omega\right)=-\frac{i}{2}\left(\Gamma_{L}+\Gamma_{R}\right)=-\frac{i\Gamma}{2}$.

Next, we look at the lesser GF's from equation (\ref{eq:steadystateGF}):
\begin{align}
G_\sigma^{<}\left(\omega\right)= & g_\sigma^{(0)<}+g_\sigma^{(0)r}\Sigma^{r}G_\sigma^{<}+g_\sigma^{(0)r}\Sigma^{<}G_\sigma^{a}+g_\sigma^{(0)<}\Sigma^{a}G_\sigma^{a}+g_\sigma^{(0)r}UG_\sigma^{\left(2\right)<}+g_\sigma^{(0)<}UG_\sigma^{\left(2\right)a}\\
G_\sigma^{\left(2\right)<}\left(\omega\right)= & g_\sigma^{\left(U\right)<}\left\langle n_{\bar{\sigma}}\right\rangle +g_\sigma^{\left(U\right)r}\Sigma^{r}G_\sigma^{\left(2\right)<}+g_\sigma^{\left(U\right)r}\Sigma^{<}G_\sigma^{\left(2\right)a}+g_\sigma^{\left(U\right)<}\Sigma^{a}G_\sigma^{\left(2\right)a}\nonumber \\
 & +g_\sigma^{\left(U\right)r}U\Gamma F_\sigma^{\left({\rm corr}\right)<}+g_\sigma^{\left(U\right)<}U\Gamma F_\sigma^{\left({\rm corr}\right)a}.\nonumber 
\end{align}
Terms of the form $g^{<}\left(1-g^{r}\Sigma^{r}\right)^{-1}$ vanish \cite{Haug2008} and we are left with:
\begin{align}
G_\sigma^{<}\left(\omega\right) & =G_\sigma^{(0)r}\Sigma^{<}G_\sigma^{a}+G_\sigma^{(0)r}UG_\sigma^{\left(2\right)<}\\
G_\sigma^{\left(2\right)<}\left(\omega\right) & =G_\sigma^{(0')r}\Sigma^{<}G_\sigma^{\left(2\right)a}+G_\sigma^{(0')r}U\Gamma F_\sigma^{\left({\rm corr}\right)<}.\nonumber 
\end{align}
By using $G^{a}=\left(G^{r}\right)^{*}$ and equation (\ref{eq:G2r}), we find
\begin{equation}
G_{\sigma}^{<}-G_\sigma^{(0)<}=U\left\langle n_{\bar{\sigma}}\right\rangle G_\sigma^{(0)r}\Sigma^{<}\left(G_\sigma^{(0')r}\right)^{*}\left(\left(G_\sigma^{(0)r}\right)^{*}+G_\sigma^{(0')r}\right)+G_{\sigma}^{\left({\rm corr}\right)<},
\end{equation}
where the lesser GF for the noninteracting model is given by
\begin{equation}
G_\sigma^{(0)<}=G_\sigma^{(0)r}\Sigma^{<}\left(G_{\sigma}^{(0)r}\right)^{*}\xrightarrow{{\rm WBL}}\frac{i}{2}\Gamma\frac{f\left(\omega-\mu_{L}\right)+f\left(\omega-\mu_{R}\right)}{\left(\omega+\frac{U}{2}\right)^{2}+\left(\frac{\Gamma}{2}\right)^{2}},
\end{equation}
and we defined 
\begin{align}
G_{\sigma}^{\left({\rm corr}\right)<}= & G_{\sigma}^{(0)r}\Sigma^{<}G_{\sigma}^{\left({\rm corr}\right)r*}+G_{\sigma}^{(0)r}UG_{\sigma}^{(0')r}\Sigma^{<}G_{\sigma}^{(0')r*}U\Gamma F^{\left({\rm corr}\right)r*}+G_{\sigma}^{(0)r}UG_{\sigma}^{(0)r}U\Gamma F^{\left({\rm corr}\right)<}\\
= & U^{2}\Gamma G_{\sigma}^{(0)r}\Sigma^{<}G_{\sigma}^{(0')r*}\left(G_{\sigma}^{(0)r*}+G_{\sigma}^{(0')r}\right)F^{\left({\rm corr}\right)r*}+U^{2}\Gamma G_{\sigma}^{(0)r}G_{\sigma}^{(0')r}F^{\left({\rm corr}\right)<}.\nonumber 
\end{align}
After simplifying and taking the wide-band limit for the lesser self
energy,
\beq
 \Sigma^{<}\left(\omega\right)=i\left(\Gamma_{L}f\left(\omega-\mu_{L}\right)+\Gamma_{R}f\left(\omega-\mu_{R}\right)\right)=\frac{i\Gamma}{2}\left(f\left(\omega-\mu_{L}\right)+f\left(\omega-\mu_{R}\right)\right),
 \eeq
the difference between the interacting and noninteracting GFs takes the form:
\begin{equation}
G_{\sigma}^{<}-G_{\sigma}^{(0)<}=i\Gamma U\left\langle n_{\bar{\sigma}}\right\rangle \frac{\omega\left(f\left(\omega-\mu_{L}\right)+f\left(\omega-\mu_{R}\right)\right)}{\left(\omega^{2}+\left(\frac{\Gamma}{2}\right)^{2}+\left(\frac{U}{2}\right)^{2}\right)^{2}+\left(\frac{\Gamma U}{2}\right)^{2}}+G_{\sigma}^{\left({\rm corr}\right)<}
\end{equation}
and the correction term is
\begin{align}
G_{\sigma}^{\left({\rm corr}\right)<}= & U^{2}\Gamma G_{\sigma}^{(0)r}\Sigma^{<}G_{\sigma}^{(0')r*}\left(G_{\sigma}^{(0)r*}+G_{\sigma}^{(0')r}\right)F^{\left({\rm corr}\right)r*}+U^{2}\Gamma G_{\sigma}^{(0)r}G_{\sigma}^{(0')r}F^{\left({\rm corr}\right)<},\\
= & U^{2}\Gamma\frac{i\Gamma\omega\left(f\left(\omega-\mu_{L}\right)+f\left(\omega-\mu_{R}\right)\right)}{\left(\omega^{2}+\left(\frac{\Gamma}{2}\right)^{2}+\left(\frac{U}{2}\right)^{2}\right)^{2}+\left(\frac{\Gamma U}{2}\right)^{2}}F^{\left({\rm corr}\right)r*}+U^{2}\Gamma\frac{1}{\left(\omega-i\frac{\Gamma}{2}\right)^{2}-\left(\frac{U}{2}\right)^{2}}F^{\left({\rm corr}\right)<}.\nonumber 
\end{align}

To recap, we have solved the steady state Green's functions for the symmetric Anderson impurity model, $G_{\sigma}^{r,<}\left(\omega\right)$, where $U=-2\varepsilon_\uparrow=-2\varepsilon_\downarrow$. These are expressed in terms of the noninteracting model with $\varepsilon=\mp U/2$, described by the Green's functions $G_{\sigma}^{(0)r,<} \left(\omega\right)$ and $G_{\sigma}^{(0')r,<} \left(\omega\right)$, respectively. Correction terms that stem from the closure of the equations of
motion, $F^{\left({\rm corr}\right)r,<}\left(\omega\right)$, are also included, to give:
\begin{align}
G_{\sigma}^{r}\left(\omega\right)-G_{\sigma}^{(0)r}\left(\omega\right) & =UG_{\sigma}^{(0)r}\left(\omega\right)G_{\sigma}^{(0')r}\left(\omega\right)\left\langle n_{\bar{\sigma}}\right\rangle +G_{\sigma}^{\left({\rm corr}\right)r}\left(\omega\right),\\ \nonumber
G_{\sigma}^{<}\left(\omega\right)-G_{\sigma}^{(0)<}\left(\omega\right) & =U\left\langle n_{\bar{\sigma}}\right\rangle G_{\sigma}^{(0)r}\left(\omega\right)\Sigma^{<}\left(\omega\right)\left(G_{\sigma}^{(0')r}\left(\omega\right)\right)^{*}\left(\left(G_{\sigma}^{(0)r}\left(\omega\right)\right)^{*}+G_{\sigma}^{(0')r}\left(\omega\right)\right)+G_{\sigma}^{\left({\rm corr}\right)<}\left(\omega\right),
\end{align}
where
\begin{align}
G_{\sigma}^{\left({\rm corr}\right)r}\left(\omega\right) & =U^{2}\Gamma G_{\sigma}^{(0)r}\left(\omega\right)G_{\sigma}^{(0')r}\left(\omega\right)F_{\sigma}^{\left({\rm corr}\right)r}\left(\omega\right),\\ \nonumber
G_{\sigma}^{\left({\rm corr}\right)<}\left(\omega\right) & =U^{2}\Gamma G_{\sigma}^{(0)r}\left(\omega\right)\Sigma^{<}\left(\omega\right)\left(G_{\sigma}^{(0')r}\left(\omega\right)\right)^{*}\left(\left(G_{\sigma}^{(0)r}\left(\omega\right)\right)^{*}+G_{\sigma}^{(0')r}\left(\omega\right)\right)\left(F^{\left({\rm corr}\right)r}\left(\omega\right)\right)^{*}\\ 
 & +U^{2}\Gamma G_{\sigma}^{(0)r}\left(\omega\right)G_{\sigma}^{(0')r}\left(\omega\right)F^{\left({\rm corr}\right)<}\left(\omega\right).\nonumber 
\end{align}
More explicitly, in the wide-band limit, the equations take the form
\begin{align}
G_{\sigma}^{r}\left(\omega\right)-G_{\sigma}^{(0)r}\left(\omega\right) & =\frac{U\left\langle n_{\bar{\sigma}}\right\rangle }{\left(\omega-i\frac{\Gamma}{2}\right)^{2}-\left(\frac{U}{2}\right)^{2}}+G_{\sigma}^{\left({\rm corr}\right)r}\left(\omega\right),\\ \nonumber
G_{\sigma}^{<}\left(\omega\right)-G_{\sigma}^{(0)<}\left(\omega\right) & =i\Gamma U\left\langle n_{\bar{\sigma}}\right\rangle \frac{\omega\left(f\left(\omega-\mu_{L}\right)+f\left(\omega-\mu_{R}\right)\right)}{\left(\omega^{2}+\left(\frac{\Gamma}{2}\right)^{2}+\left(\frac{U}{2}\right)^{2}\right)^{2}+\left(\frac{\Gamma U}{2}\right)^{2}}+G_{\sigma}^{\left({\rm corr}\right)<}\left(\omega\right).
\end{align}

\subsection{Steady state current calculation}
First recall that in the symmetric case of the Anderson model,
\begin{equation}
I_{\uparrow}=I_{\downarrow}=\frac{I}{2},
\end{equation}
therefore, we will look at the current of one spin, where the full current will be twice the former. The particle current from the $\ell$ lead in steady state for both the symmetric and the noninteracting models, is given by:
\begin{equation}
I_{\ell\sigma}=2\int\frac{d\omega}{2\pi}\left[i\Sigma_\ell^{<}\left(\omega\right)\Im\left\{ G_\sigma^{r}\left(\omega\right)\right\} +\Sigma_{\ell}^{r}\left(\omega\right)G_\sigma^{<}\left(\omega\right)\right],
\end{equation}
where the electric current will be given by multiplying by the charge of the electron, $e$.
The difference between the Anderson current (for one of the
spins) and the  noninteracting model current is therefore given by:
\begin{equation}
I_{\ell\sigma}^{\left(U=-2\varepsilon\right)}-I_{\ell\sigma}^{\left(U=0\right)}=2\int\frac{d\omega}{2\pi}\left[i\Sigma_{\ell}^{<}\Im\left\{ G_{\sigma}^{r}-G_{\sigma}^{(0)r}\right\} +\Sigma_{\ell}^{r}\left(G_{\sigma}^{<}-G_{\sigma}^{(0)<}\right)\right].
\end{equation}
Using the results above, we find:
\begin{equation}
I_{\ell\sigma}^{\left(U=-2\varepsilon\right)}-I_{\ell\sigma}^{\left(U=0\right)}=\frac{\Gamma^{2}}{2}U\left\langle n_{\bar{\sigma}}\right\rangle \int\frac{d\omega}{2\pi}\left[\frac{\omega\left(f\left(\omega-\mu_{L}\right)-f\left(\omega-\mu_{R}\right)\right)}{\omega^{4}+2\omega^{2}\left(\left(\frac{\Gamma}{2}\right)^{2}-\left(\frac{U}{2}\right)^{2}\right)+\left(\left(\frac{\Gamma}{2}\right)^{2}+\left(\frac{U}{2}\right)^{2}\right)^{2}}\right]+I_{\ell\sigma}^{\left({\rm corr}\right)}
\end{equation}
where we defined the contribution to the current as a result of the
corrections $G_\sigma^{\left({\rm corr}\right)}$ due to higher order GFs which were neglected in the closure, as:
\begin{align}
I_{\ell	\sigma}^{\left({\rm corr}\right)}= & -\Gamma\int\frac{d\omega}{2\pi}\left[f\left(\omega-\mu_{L}\right)\Im\left\{ G_\sigma^{\left({\rm corr}\right)r}\right\} +iG_\sigma^{\left({\rm corr}\right)<}\right].
\end{align}
Since $G^{\left({\rm corr}\right)}\sim\Gamma U^{2}$ and $\Gamma^{2}U^{2}$,
the correction to the current will contribute as $I^{\left({\rm corr}\right)}\sim\Gamma^{2}U^{2}$
and $\Gamma^{3}U^{2}$.

When the electric bias is centred around the Fermi level,
i.e. $\mu_{L}=-\mu_{R}=\mu$, the difference between the
noninteracting model and symmetric-Anderson model currents
is zero up to the high order correction due to the closure.
\begin{align}
I_{\ell\sigma}^{\left(U=-2\varepsilon\right)}-I_{\ell\sigma}^{\left(U=0\right)} & =\frac{\Gamma^{2}}{2}U\left\langle n_{\bar{\sigma}}\right\rangle \int_{-\infty}^{\infty}\frac{d\omega}{2\pi}\frac{\omega\left(f\left(\omega-\mu_{L}\right)-f\left(\omega-\mu_{R}\right)\right)}{\omega^{4}+2\omega^{2}\left(\left(\frac{\Gamma}{2}\right)^{2}-\left(\frac{U}{2}\right)^{2}\right)+\left(\left(\frac{\Gamma}{2}\right)^{2}+\left(\frac{U}{2}\right)^{2}\right)^{2}}+I_{\ell	\sigma}^{\left({\rm corr}\right)}\nonumber \\
 & =\frac{\Gamma^{2}}{2}U\left\langle n_{\bar{\sigma}}\right\rangle \int_{-\infty}^{\infty}\frac{d\omega}{2\pi}\underbrace{\omega}_{{\rm odd}}\underbrace{\frac{f\left(\omega-\mu\right)-f\left(\omega+\mu\right)}{\omega^{4}+2\omega^{2}\left(\left(\frac{\Gamma}{2}\right)^{2}-\left(\frac{U}{2}\right)^{2}\right)+\left(\left(\frac{\Gamma}{2}\right)^{2}+\left(\frac{U}{2}\right)^{2}\right)^{2}}}_{{\rm even}}+I_{\ell\sigma}^{\left({\rm corr}\right)}\nonumber \\
 & =0+I_{\ell\sigma}^{\left({\rm corr}\right)}.
\end{align}
This results shows that within the closure specified above, which captures the Coulomb blockade, at the symmetric point, the noninteracting and interacting currents are equal. Furthermore, the leading corrections to this result scale as $\Gamma^2 U^2$.

\section{Comparing the NEGF and ME approaches}
To justify the validity of the results obtained using the ME approach we assess the validity of the ME by comparing the results to the NEGF which becomes exact for the noninteracting system.
Fig. \ref{fig:figS1} demonstrates the regime of validity of the ME. We plot the relative error in the steady-state current for the noninteracting system, $\vert I^{exact} -I^{ME} \vert /I^{exact} $. 
The exact current is calculated using the NEGF approach which is exact for $U=0$.
We find that when  $T,\varepsilon\gtrsim 10\Gamma$  the relative error is $\sim1\%$.
\begin{figure}[!htb]
\minipage{0.45\textwidth}
  \includegraphics[width=\linewidth]{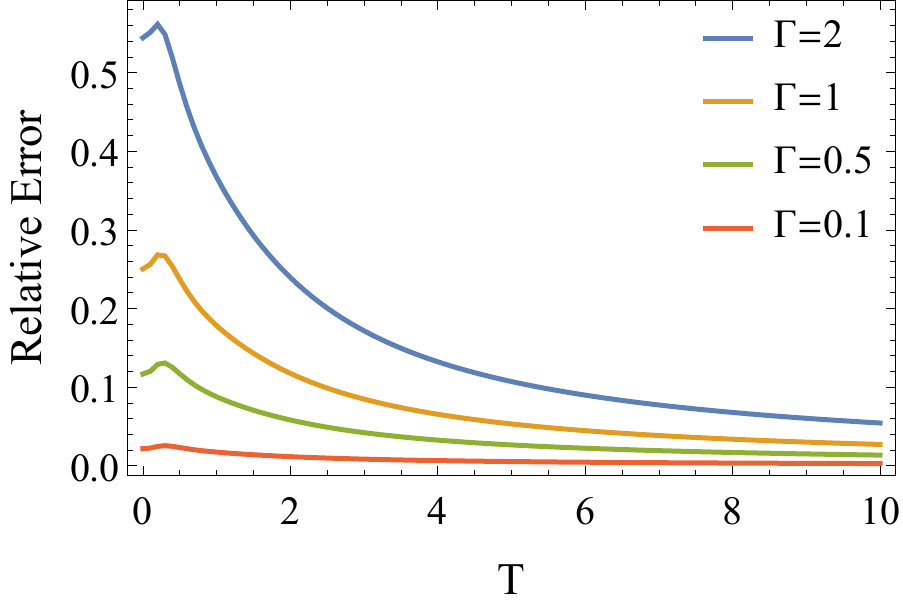}
\endminipage\hfill
\minipage{0.45\textwidth}
  \includegraphics[width=\linewidth]{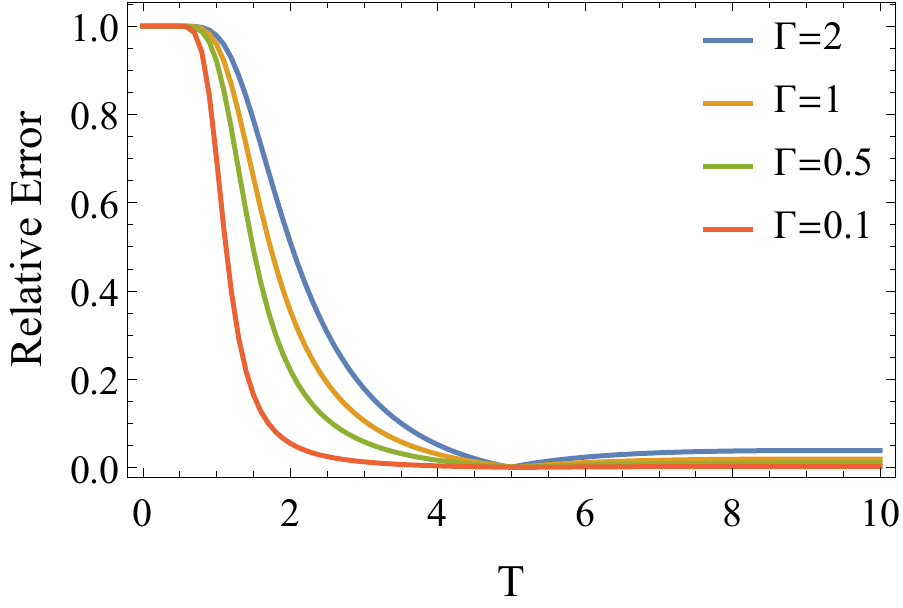}
\endminipage\
\caption{The relative error in the current for a noninteracting system calculated using the ME, as function of the leads' temperature, for different coupling strengths $\Gamma$. In (a) $\varepsilon=-1$, and in (b) $\varepsilon=-10$. In both panels: $\varepsilon_f=0$, $V=4$, $\Gamma_L=\Gamma_R=\Gamma/2$ and $U=0$.  
}
\label{fig:figS1}
\end{figure}
\par
 
In Fig. \ref{fig:figS2} we plot the steady state current for the interacting system using the NEGF and its absolute value difference from the current obtained from the ME,  $|I^{NEGF}-I^{ME}|$. At sufficiently high temperatures ($T\sim 10\Gamma$) the currents calculated from NEGF and the ME are in good agreement.

\begin{figure}[H] 
  \begin{minipage}{0.4\textwidth}
    \centering
    \includegraphics[width=\linewidth]{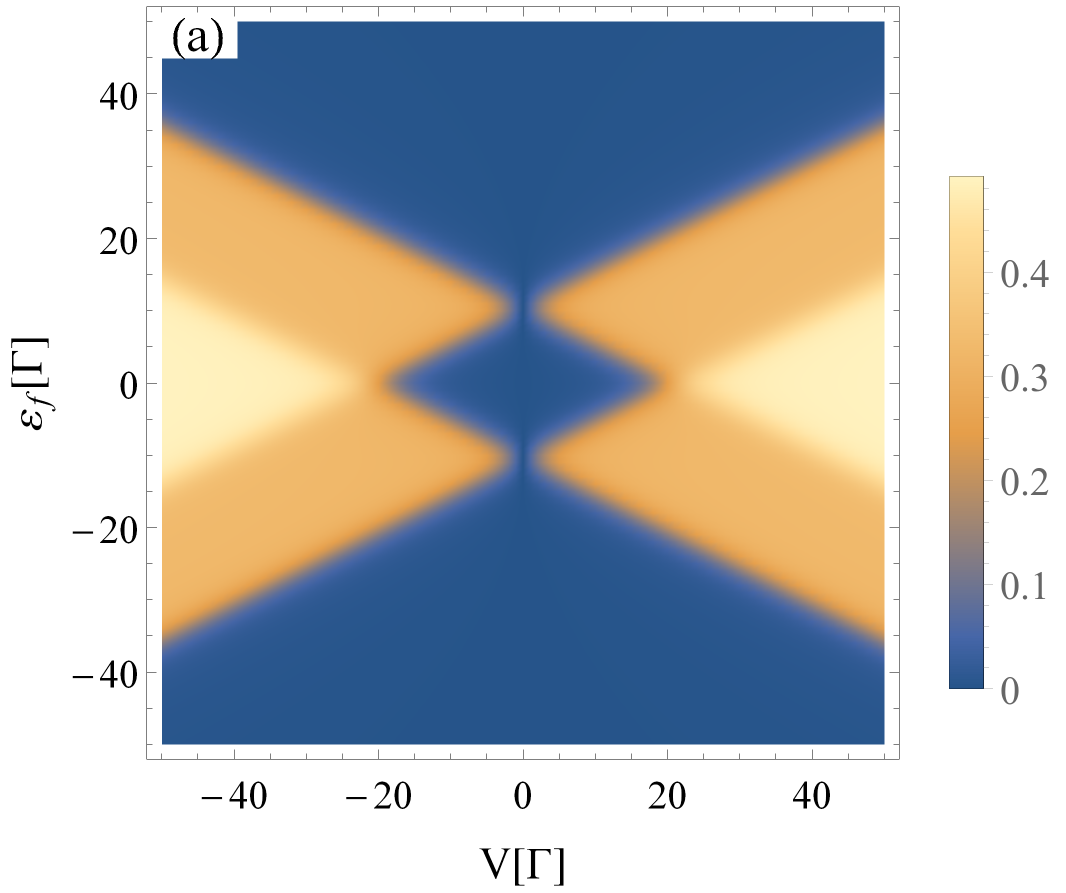} 
  \end{minipage}
  \begin{minipage}{0.4\textwidth}
    \centering
    \includegraphics[width=\linewidth]{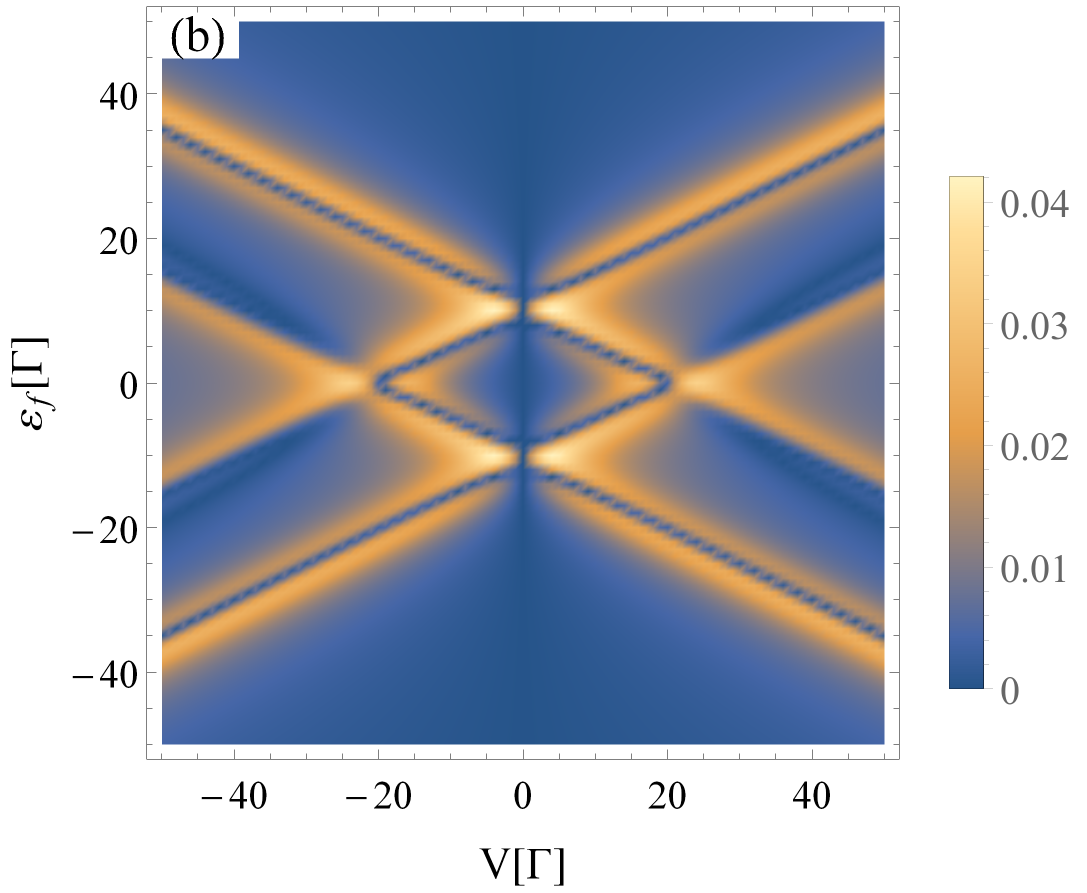} 
  \end{minipage} 
  \begin{minipage}{0.4\textwidth}
    \centering
    \includegraphics[width=\linewidth]{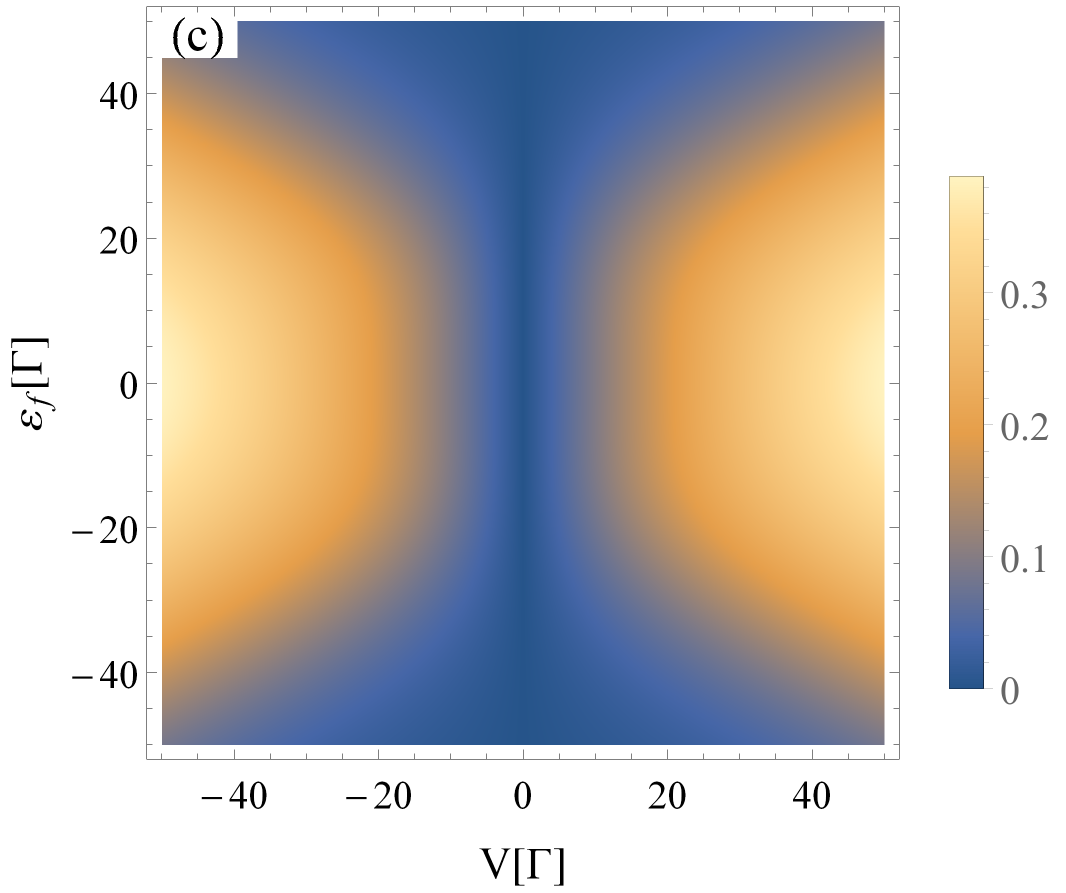} 
  \end{minipage}
  \begin{minipage}{0.4\textwidth}
    \centering
    \includegraphics[width=\linewidth]{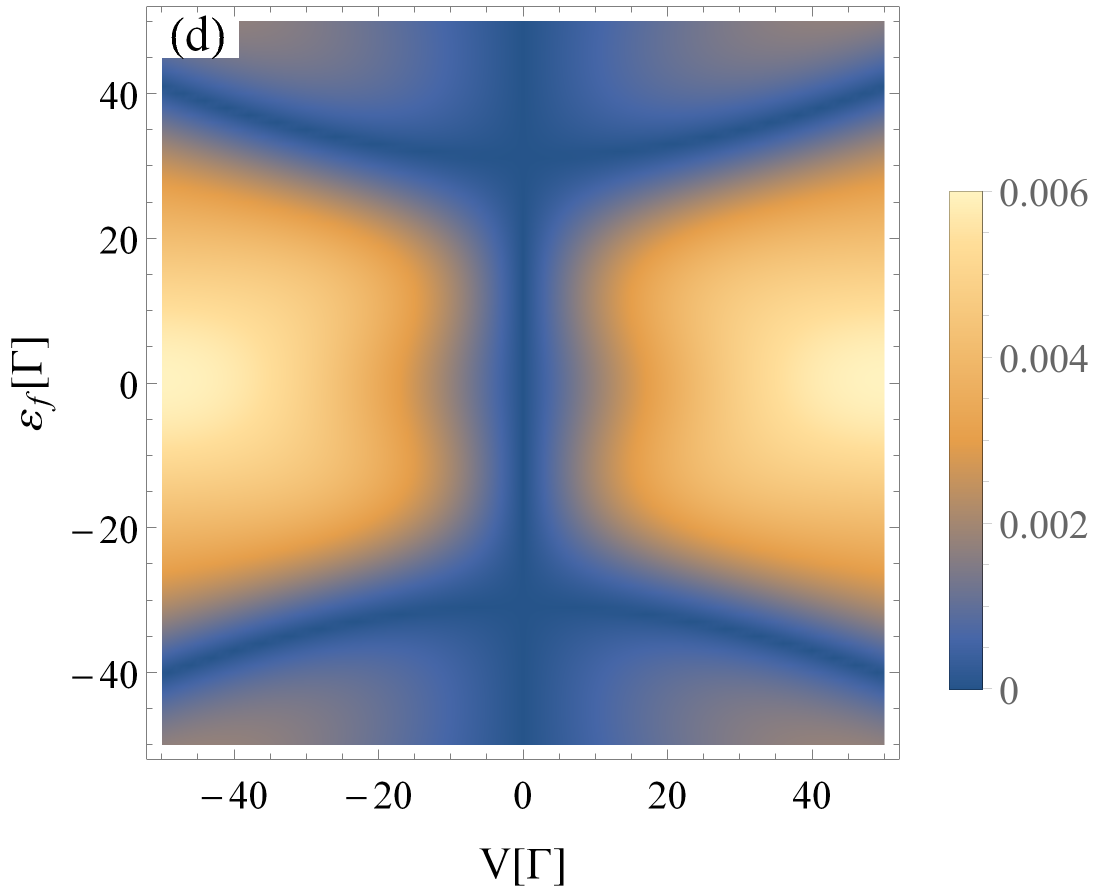} 
  \end{minipage} 
  \caption{Left panels: the magnitude of the steady-state current in the symmetric system calculated using NEGF.  Right panels: the absolute value difference between the currents calculated form NEGF and the ME. All panels are plotted as function of the Fermi-energy $\varepsilon_f$ and the bias voltage $V=0$. In the top panels $T_L=T_R=\Gamma$, and in the bottom $T_L=T_R=10\Gamma$. In all panels: $\varepsilon=-10\Gamma $, $U=-2\varepsilon$, and $\Gamma_L=\Gamma_r=\Gamma/2$. Note the different scales.}
\label{fig:figS2} 
\end{figure}   

\end{suppinfo}


\begin{mcitethebibliography}{56}
\providecommand*\natexlab[1]{#1}
\providecommand*\mciteSetBstSublistMode[1]{}
\providecommand*\mciteSetBstMaxWidthForm[2]{}
\providecommand*\mciteBstWouldAddEndPuncttrue
  {\def\EndOfBibitem{\unskip.}}
\providecommand*\mciteBstWouldAddEndPunctfalse
  {\let\EndOfBibitem\relax}
\providecommand*\mciteSetBstMidEndSepPunct[3]{}
\providecommand*\mciteSetBstSublistLabelBeginEnd[3]{}
\providecommand*\EndOfBibitem{}
\mciteSetBstSublistMode{f}
\mciteSetBstMaxWidthForm{subitem}{(\alph{mcitesubitemcount})}
\mciteSetBstSublistLabelBeginEnd
  {\mcitemaxwidthsubitemform\space}
  {\relax}
  {\relax}

\bibitem[Anderson(1961)]{anderson1961}
Anderson,~P.~W. Localized Magnetic States in Metals. \emph{Phys. Rev.}
  \textbf{1961}, \emph{124}, 41--53\relax
\mciteBstWouldAddEndPuncttrue
\mciteSetBstMidEndSepPunct{\mcitedefaultmidpunct}
{\mcitedefaultendpunct}{\mcitedefaultseppunct}\relax
\EndOfBibitem
\bibitem[Goldhaber-Gordon \latin{et~al.}(1998)Goldhaber-Gordon, Shtrikman,
  Mahalu, Abusch-Magder, Meirav, and Kastner]{Kastner1998}
Goldhaber-Gordon,~D.; Shtrikman,~H.; Mahalu,~D.; Abusch-Magder,~D.; Meirav,~U.;
  Kastner,~M.~A. Kondo Effect in a Single-Electron Transistor. \emph{Nature
  (London)} \textbf{1998}, \emph{391}, 156--159\relax
\mciteBstWouldAddEndPuncttrue
\mciteSetBstMidEndSepPunct{\mcitedefaultmidpunct}
{\mcitedefaultendpunct}{\mcitedefaultseppunct}\relax
\EndOfBibitem
\bibitem[Hanson \latin{et~al.}(2007)Hanson, Kouwenhoven, Petta, Tarucha, and
  Vandersypen]{hanson2007spins}
Hanson,~R.; Kouwenhoven,~L.; Petta,~J.; Tarucha,~S.; Vandersypen,~L. Spins in
  Few-Electron Quantum Dots. \emph{Rev. Mod. Phys.} \textbf{2007}, \emph{79},
  1217--1265\relax
\mciteBstWouldAddEndPuncttrue
\mciteSetBstMidEndSepPunct{\mcitedefaultmidpunct}
{\mcitedefaultendpunct}{\mcitedefaultseppunct}\relax
\EndOfBibitem
\bibitem[Nitzan and Ratner(2003)Nitzan, and Ratner]{Nitzan2003}
Nitzan,~A.; Ratner,~M.~A. Electron Transport in Molecular Wire Junctions.
  \emph{Science} \textbf{2003}, \emph{300}, 1384--1389\relax
\mciteBstWouldAddEndPuncttrue
\mciteSetBstMidEndSepPunct{\mcitedefaultmidpunct}
{\mcitedefaultendpunct}{\mcitedefaultseppunct}\relax
\EndOfBibitem
\bibitem[Heath and Ratner(2003)Heath, and Ratner]{Heath2003}
Heath,~J.~R.; Ratner,~M.~A. Molecular Electronics. \emph{Phys. Today}
  \textbf{2003}, \emph{56}, 43--49\relax
\mciteBstWouldAddEndPuncttrue
\mciteSetBstMidEndSepPunct{\mcitedefaultmidpunct}
{\mcitedefaultendpunct}{\mcitedefaultseppunct}\relax
\EndOfBibitem
\bibitem[Beenakker(1991)]{Beenakker1991}
Beenakker,~C. W.~J. Theory of Coulomb-Blockade Oscillations in the Conductance
  of a Quantum Dot. \emph{Phys. Rev. B} \textbf{1991}, \emph{44},
  1646--1656\relax
\mciteBstWouldAddEndPuncttrue
\mciteSetBstMidEndSepPunct{\mcitedefaultmidpunct}
{\mcitedefaultendpunct}{\mcitedefaultseppunct}\relax
\EndOfBibitem
\bibitem[Koch \latin{et~al.}(2006)Koch, Raikh, and Von~Oppen]{koch2006}
Koch,~J.; Raikh,~M.~E.; Von~Oppen,~F. Pair Tunneling Through Single Molecules.
  \emph{Phys. Rev. Lett.} \textbf{2006}, \emph{96}, 056803\relax
\mciteBstWouldAddEndPuncttrue
\mciteSetBstMidEndSepPunct{\mcitedefaultmidpunct}
{\mcitedefaultendpunct}{\mcitedefaultseppunct}\relax
\EndOfBibitem
\bibitem[Kondo(1964)]{kondo1964}
Kondo,~J. Resistance Minimum in Dilute Magnetic Alloys. \emph{Prog. Theor.
  Phys.} \textbf{1964}, \emph{32}, 37--49\relax
\mciteBstWouldAddEndPuncttrue
\mciteSetBstMidEndSepPunct{\mcitedefaultmidpunct}
{\mcitedefaultendpunct}{\mcitedefaultseppunct}\relax
\EndOfBibitem
\bibitem[Schrieffer and Wolff(1966)Schrieffer, and Wolff]{schrieffer1966}
Schrieffer,~J.~R.; Wolff,~P.~A. Relation Between the Anderson and Kondo
  Hamiltonians. \emph{Phys. Rev.} \textbf{1966}, \emph{149}, 491--492\relax
\mciteBstWouldAddEndPuncttrue
\mciteSetBstMidEndSepPunct{\mcitedefaultmidpunct}
{\mcitedefaultendpunct}{\mcitedefaultseppunct}\relax
\EndOfBibitem
\bibitem[Datta(1990)]{Datta1990}
Datta,~S. A Simple Kinetic-Equation for Steady-State Quantum Transport.
  \emph{J. Phys. C} \textbf{1990}, \emph{2}, 8023--8052\relax
\mciteBstWouldAddEndPuncttrue
\mciteSetBstMidEndSepPunct{\mcitedefaultmidpunct}
{\mcitedefaultendpunct}{\mcitedefaultseppunct}\relax
\EndOfBibitem
\bibitem[Harbola \latin{et~al.}(2006)Harbola, Esposito, and
  Mukamel]{Mukamel2006}
Harbola,~U.; Esposito,~M.; Mukamel,~S. Quantum Master Equation for Electron
  Transport Through Quantum Dots and Single Molecules. \emph{Phys. Rev. B}
  \textbf{2006}, \emph{74}, 235309\relax
\mciteBstWouldAddEndPuncttrue
\mciteSetBstMidEndSepPunct{\mcitedefaultmidpunct}
{\mcitedefaultendpunct}{\mcitedefaultseppunct}\relax
\EndOfBibitem
\bibitem[Leijnse and Wegewijs(2008)Leijnse, and Wegewijs]{Leijnse2008}
Leijnse,~M.; Wegewijs,~M.~R. Kinetic Equations for Transport Through
  Single-Molecule Transistors. \emph{Phys. Rev. B} \textbf{2008}, \emph{78},
  235424\relax
\mciteBstWouldAddEndPuncttrue
\mciteSetBstMidEndSepPunct{\mcitedefaultmidpunct}
{\mcitedefaultendpunct}{\mcitedefaultseppunct}\relax
\EndOfBibitem
\bibitem[Esposito and Galperin(2009)Esposito, and Galperin]{Esposito2009}
Esposito,~M.; Galperin,~M. Transport in Molecular States Language: Generalized
  Quantum Master Equation Approach. \emph{Phys. Rev. B} \textbf{2009},
  \emph{79}, 205303\relax
\mciteBstWouldAddEndPuncttrue
\mciteSetBstMidEndSepPunct{\mcitedefaultmidpunct}
{\mcitedefaultendpunct}{\mcitedefaultseppunct}\relax
\EndOfBibitem
\bibitem[Esposito and Galperin(2010)Esposito, and Galperin]{Esposito2010}
Esposito,~M.; Galperin,~M. Self-Consistent Quantum Master Equation Approach to
  Molecular Transport. \emph{J. Phys. Chem. C} \textbf{2010}, \emph{114},
  20362--20369\relax
\mciteBstWouldAddEndPuncttrue
\mciteSetBstMidEndSepPunct{\mcitedefaultmidpunct}
{\mcitedefaultendpunct}{\mcitedefaultseppunct}\relax
\EndOfBibitem
\bibitem[Dou \latin{et~al.}(2015)Dou, Nitzan, and Subotnik]{dou2015}
Dou,~W.; Nitzan,~A.; Subotnik,~J.~E. Surface Hopping with a Manifold of
  Electronic States. III. Transients, Broadening, and the Marcus Picture.
  \emph{J. Chem. Phys.} \textbf{2015}, \emph{142}, 234106\relax
\mciteBstWouldAddEndPuncttrue
\mciteSetBstMidEndSepPunct{\mcitedefaultmidpunct}
{\mcitedefaultendpunct}{\mcitedefaultseppunct}\relax
\EndOfBibitem
\bibitem[Dorda \latin{et~al.}(2015)Dorda, Ganahl, Evertz, Von Der~Linden, and
  Arrigoni]{dorda2015}
Dorda,~A.; Ganahl,~M.; Evertz,~H.~G.; Von Der~Linden,~W.; Arrigoni,~E.
  Auxiliary Master Equation Approach within Matrix Product States: Spectral
  Properties of the Nonequilibrium Anderson Impurity Model. \emph{Phys. Rev. B}
  \textbf{2015}, \emph{92}, 125145\relax
\mciteBstWouldAddEndPuncttrue
\mciteSetBstMidEndSepPunct{\mcitedefaultmidpunct}
{\mcitedefaultendpunct}{\mcitedefaultseppunct}\relax
\EndOfBibitem
\bibitem[Hettler \latin{et~al.}(1998)Hettler, Kroha, and
  Hershfield]{hettler1998}
Hettler,~M.~H.; Kroha,~J.; Hershfield,~S. Nonequilibrium Dynamics of the
  Anderson Impurity Model. \emph{Phys. Rev. B} \textbf{1998}, \emph{58},
  5649--5664\relax
\mciteBstWouldAddEndPuncttrue
\mciteSetBstMidEndSepPunct{\mcitedefaultmidpunct}
{\mcitedefaultendpunct}{\mcitedefaultseppunct}\relax
\EndOfBibitem
\bibitem[Datta(2000)]{Datta2000}
Datta,~S. Nanoscale Device Modeling: the Green's Function Method.
  \emph{Superlattices and Microstructures} \textbf{2000}, \emph{28},
  253--278\relax
\mciteBstWouldAddEndPuncttrue
\mciteSetBstMidEndSepPunct{\mcitedefaultmidpunct}
{\mcitedefaultendpunct}{\mcitedefaultseppunct}\relax
\EndOfBibitem
\bibitem[Xue \latin{et~al.}(2002)Xue, Datta, and Ratner]{Xue2002}
Xue,~Y.~Q.; Datta,~S.; Ratner,~M.~A. First-Principles Based Matrix Green's
  Function Approach to Molecular Electronic Devices: General Formalism.
  \emph{Chem. Phys.} \textbf{2002}, \emph{281}, 151--170\relax
\mciteBstWouldAddEndPuncttrue
\mciteSetBstMidEndSepPunct{\mcitedefaultmidpunct}
{\mcitedefaultendpunct}{\mcitedefaultseppunct}\relax
\EndOfBibitem
\bibitem[Galperin \latin{et~al.}(2007)Galperin, Ratner, and Nitzan]{Galperin07}
Galperin,~M.; Ratner,~M.~A.; Nitzan,~A. Molecular Transport Junctions:
  Vibrational Effects. \emph{J. Phys.: Condens. Matter} \textbf{2007},
  \emph{19}, 103201\relax
\mciteBstWouldAddEndPuncttrue
\mciteSetBstMidEndSepPunct{\mcitedefaultmidpunct}
{\mcitedefaultendpunct}{\mcitedefaultseppunct}\relax
\EndOfBibitem
\bibitem[Galperin \latin{et~al.}(2007)Galperin, Nitzan, and
  Ratner]{galperin07b}
Galperin,~M.; Nitzan,~A.; Ratner,~M.~A. Inelastic Effects in Molecular
  Junctions in the Coulomb and Kondo Regimes: Nonequilibrium Equation-of-Motion
  Approach. \emph{Phys. Rev. B} \textbf{2007}, \emph{76}, 035301\relax
\mciteBstWouldAddEndPuncttrue
\mciteSetBstMidEndSepPunct{\mcitedefaultmidpunct}
{\mcitedefaultendpunct}{\mcitedefaultseppunct}\relax
\EndOfBibitem
\bibitem[Haug and Jauho(2008)Haug, and Jauho]{Haug2008}
Haug,~H.; Jauho,~A.-P. \emph{Quantum Kinetics in Transport and Optics of
  Semiconductors}, 2nd ed.; Springer series in solid-state sciences,; Springer:
  Berlin ; New York, 2008\relax
\mciteBstWouldAddEndPuncttrue
\mciteSetBstMidEndSepPunct{\mcitedefaultmidpunct}
{\mcitedefaultendpunct}{\mcitedefaultseppunct}\relax
\EndOfBibitem
\bibitem[Chen \latin{et~al.}(2017)Chen, Gao, and Galperin]{galperin17}
Chen,~F.; Gao,~Y.; Galperin,~M. Molecular Heat Engines: Quantum Coherence
  Effects. \emph{Entropy} \textbf{2017}, \emph{19}, 472\relax
\mciteBstWouldAddEndPuncttrue
\mciteSetBstMidEndSepPunct{\mcitedefaultmidpunct}
{\mcitedefaultendpunct}{\mcitedefaultseppunct}\relax
\EndOfBibitem
\bibitem[Stefanucci and Leeuwen(2013)Stefanucci, and
  Leeuwen]{stefanucci_nonequilibrium_2013}
Stefanucci,~G.; Leeuwen,~R.~v. \emph{Nonequilibrium Many-Body Theory of Quantum
  Systems: A Modern Introduction}; Cambridge University Press, 2013\relax
\mciteBstWouldAddEndPuncttrue
\mciteSetBstMidEndSepPunct{\mcitedefaultmidpunct}
{\mcitedefaultendpunct}{\mcitedefaultseppunct}\relax
\EndOfBibitem
\bibitem[Li \latin{et~al.}(2013)Li, Levy, Swenson, Rabani, and Miller]{li2013}
Li,~B.; Levy,~T.~J.; Swenson,~D.~W.; Rabani,~E.; Miller,~W.~H. A Cartesian
  Quasi-Classical Model to Nonequilibrium Quantum Transport: The Anderson
  Impurity Model. \emph{J. Chem. Phys.} \textbf{2013}, \emph{138}, 104110\relax
\mciteBstWouldAddEndPuncttrue
\mciteSetBstMidEndSepPunct{\mcitedefaultmidpunct}
{\mcitedefaultendpunct}{\mcitedefaultseppunct}\relax
\EndOfBibitem
\bibitem[Li \latin{et~al.}(2014)Li, Miller, Levy, and Rabani]{li2014}
Li,~B.; Miller,~W.~H.; Levy,~T.~J.; Rabani,~E. Classical Mapping for Hubbard
  Operators: Application to the Double-Anderson Model. \emph{J. Chem. Phys.}
  \textbf{2014}, \emph{140}, 204106\relax
\mciteBstWouldAddEndPuncttrue
\mciteSetBstMidEndSepPunct{\mcitedefaultmidpunct}
{\mcitedefaultendpunct}{\mcitedefaultseppunct}\relax
\EndOfBibitem
\bibitem[M\"uhlbacher and Rabani(2008)M\"uhlbacher, and Rabani]{rabani2008}
M\"uhlbacher,~L.; Rabani,~E. Real-Time Path Integral Approach to Nonequilibrium
  Many-Body Quantum Systems. \emph{Phys. Rev. Lett.} \textbf{2008}, \emph{100},
  176403\relax
\mciteBstWouldAddEndPuncttrue
\mciteSetBstMidEndSepPunct{\mcitedefaultmidpunct}
{\mcitedefaultendpunct}{\mcitedefaultseppunct}\relax
\EndOfBibitem
\bibitem[Weiss \latin{et~al.}(2008)Weiss, Eckel, Thorwart, and
  Egger]{weiss_iterative_2008}
Weiss,~S.; Eckel,~J.; Thorwart,~M.; Egger,~R. Iterative Real-Time Path Integral
  Approach to Nonequilibrium Quantum Transport. \emph{Phys. Rev. B}
  \textbf{2008}, \emph{77}, 195316\relax
\mciteBstWouldAddEndPuncttrue
\mciteSetBstMidEndSepPunct{\mcitedefaultmidpunct}
{\mcitedefaultendpunct}{\mcitedefaultseppunct}\relax
\EndOfBibitem
\bibitem[Schir\'{o} and Fabrizio(2009)Schir\'{o}, and
  Fabrizio]{schiro_real-time_2009}
Schir\'{o},~M.; Fabrizio,~M. Real-Time Diagrammatic Monte Carlo for
  Nonequilibrium Quantum Transport. \emph{Phys. Rev. B} \textbf{2009},
  \emph{79}, 153302\relax
\mciteBstWouldAddEndPuncttrue
\mciteSetBstMidEndSepPunct{\mcitedefaultmidpunct}
{\mcitedefaultendpunct}{\mcitedefaultseppunct}\relax
\EndOfBibitem
\bibitem[Werner \latin{et~al.}(2009)Werner, Oka, and
  Millis]{werner_diagrammatic_2009}
Werner,~P.; Oka,~T.; Millis,~A.~J. Diagrammatic Monte Carlo Simulation of
  Nonequilibrium Systems. \emph{Phys. Rev. B} \textbf{2009}, \emph{79},
  035320\relax
\mciteBstWouldAddEndPuncttrue
\mciteSetBstMidEndSepPunct{\mcitedefaultmidpunct}
{\mcitedefaultendpunct}{\mcitedefaultseppunct}\relax
\EndOfBibitem
\bibitem[Gull \latin{et~al.}(2010)Gull, Reichman, and
  Millis]{gull10_bold_monte_carlo}
Gull,~E.; Reichman,~D.~R.; Millis,~A.~J. Bold-Line Diagrammatic Monte Carlo
  Method: General Formulation and Application to Expansion Around the
  Noncrossing Approximation. \emph{Phys. Rev. B} \textbf{2010}, \emph{82},
  075109\relax
\mciteBstWouldAddEndPuncttrue
\mciteSetBstMidEndSepPunct{\mcitedefaultmidpunct}
{\mcitedefaultendpunct}{\mcitedefaultseppunct}\relax
\EndOfBibitem
\bibitem[Segal \latin{et~al.}(2010)Segal, Millis, and Reichman]{Segal10}
Segal,~D.; Millis,~A.~J.; Reichman,~D.~R. Numerically Exact Path-Integral
  Simulation of Nonequilibrium Quantum Transport and Dissipation. \emph{Phys.
  Rev. B} \textbf{2010}, \emph{82}, 205323\relax
\mciteBstWouldAddEndPuncttrue
\mciteSetBstMidEndSepPunct{\mcitedefaultmidpunct}
{\mcitedefaultendpunct}{\mcitedefaultseppunct}\relax
\EndOfBibitem
\bibitem[Cohen and Rabani(2011)Cohen, and Rabani]{cohen_memory_2011}
Cohen,~G.; Rabani,~E. Memory Effects in Nonequilibrium Quantum Impurity Models.
  \emph{Phys. Rev. B} \textbf{2011}, \emph{84}, 075150\relax
\mciteBstWouldAddEndPuncttrue
\mciteSetBstMidEndSepPunct{\mcitedefaultmidpunct}
{\mcitedefaultendpunct}{\mcitedefaultseppunct}\relax
\EndOfBibitem
\bibitem[Hartle \latin{et~al.}(2013)Hartle, Cohen, Reichman, and
  Millis]{Hartle2013}
Hartle,~R.; Cohen,~G.; Reichman,~D.~R.; Millis,~A.~J. Decoherence and
  Lead-Induced Interdot Coupling in Nonequilibrium Electron Transport Through
  Interacting Quantum Dots: A Hierarchical Quantum Master Equation Approach.
  \emph{Phys. Rev. B} \textbf{2013}, \emph{88}, 235426\relax
\mciteBstWouldAddEndPuncttrue
\mciteSetBstMidEndSepPunct{\mcitedefaultmidpunct}
{\mcitedefaultendpunct}{\mcitedefaultseppunct}\relax
\EndOfBibitem
\bibitem[Cohen \latin{et~al.}(2015)Cohen, Gull, Reichman, and
  Millis]{Cohen-Gull-Reichman2015-introducing-inchworm}
Cohen,~G.; Gull,~E.; Reichman,~D.~R.; Millis,~A.~J. Taming the Dynamical Sign
  Problem in Real-Time Evolution of Quantum Many-Body Problems. \emph{Phys.
  Rev. Lett.} \textbf{2015}, \emph{115}, 266802\relax
\mciteBstWouldAddEndPuncttrue
\mciteSetBstMidEndSepPunct{\mcitedefaultmidpunct}
{\mcitedefaultendpunct}{\mcitedefaultseppunct}\relax
\EndOfBibitem
\bibitem[Schmitteckert(2004)]{schmitteckert_nonequilibrium_2004}
Schmitteckert,~P. Nonequilibrium Electron Transport Using the Density Matrix
  Renormalization Group Method. \emph{Phys. Rev. B} \textbf{2004}, \emph{70},
  121302\relax
\mciteBstWouldAddEndPuncttrue
\mciteSetBstMidEndSepPunct{\mcitedefaultmidpunct}
{\mcitedefaultendpunct}{\mcitedefaultseppunct}\relax
\EndOfBibitem
\bibitem[Anders and Schiller(2005)Anders, and Schiller]{anders_real-time_2005}
Anders,~F.~B.; Schiller,~A. {Real-Time} Dynamics in {Quantum-Impurity} Systems:
  A {Time-Dependent} Numerical {Renormalization-Group} Approach. \emph{Phys.
  Rev. Lett.} \textbf{2005}, \emph{95}, 196801\relax
\mciteBstWouldAddEndPuncttrue
\mciteSetBstMidEndSepPunct{\mcitedefaultmidpunct}
{\mcitedefaultendpunct}{\mcitedefaultseppunct}\relax
\EndOfBibitem
\bibitem[Bulla \latin{et~al.}(2008)Bulla, Costi, and
  Pruschke]{bulla2008numerical}
Bulla,~R.; Costi,~T.~A.; Pruschke,~T. Numerical Renormalization Group Method
  for Quantum Impurity Systems. \emph{Rev. Mod. Phys.} \textbf{2008},
  \emph{80}, 395\relax
\mciteBstWouldAddEndPuncttrue
\mciteSetBstMidEndSepPunct{\mcitedefaultmidpunct}
{\mcitedefaultendpunct}{\mcitedefaultseppunct}\relax
\EndOfBibitem
\bibitem[Wang and Thoss(2009)Wang, and Thoss]{Wang2009}
Wang,~H.; Thoss,~M. Numerically Exact Quantum Dynamics for Indistinguishable
  Particles: The Multilayer Multiconfiguration Time-Dependent Hartree Theory in
  Second Quantization Representation. \emph{J. Chem. Phys.} \textbf{2009},
  \emph{131}, 024114\relax
\mciteBstWouldAddEndPuncttrue
\mciteSetBstMidEndSepPunct{\mcitedefaultmidpunct}
{\mcitedefaultendpunct}{\mcitedefaultseppunct}\relax
\EndOfBibitem
\bibitem[Hofstetter and Kehrein(1999)Hofstetter, and Kehrein]{hofstetter1999}
Hofstetter,~W.; Kehrein,~S. Symmetric Anderson Impurity Model with a Narrow
  Band. \emph{Phys. Rev. B} \textbf{1999}, \emph{59}, R12732\relax
\mciteBstWouldAddEndPuncttrue
\mciteSetBstMidEndSepPunct{\mcitedefaultmidpunct}
{\mcitedefaultendpunct}{\mcitedefaultseppunct}\relax
\EndOfBibitem
\bibitem[Dickens and Logan(2001)Dickens, and Logan]{dickens2001}
Dickens,~N.~L.; Logan,~D.~E. On the Scaling Spectrum of the Anderson Impurity
  Model. \emph{J. Phys.: Condens. Matter} \textbf{2001}, \emph{13}, 4505\relax
\mciteBstWouldAddEndPuncttrue
\mciteSetBstMidEndSepPunct{\mcitedefaultmidpunct}
{\mcitedefaultendpunct}{\mcitedefaultseppunct}\relax
\EndOfBibitem
\bibitem[Motahari \latin{et~al.}(2016)Motahari, Requist, and
  Jacob]{motahari2016}
Motahari,~S.; Requist,~R.; Jacob,~D. Kondo Physics of the Anderson Impurity
  Model by Distributional Exact Diagonalization. \emph{Phys. Rev. B}
  \textbf{2016}, \emph{94}, 235133\relax
\mciteBstWouldAddEndPuncttrue
\mciteSetBstMidEndSepPunct{\mcitedefaultmidpunct}
{\mcitedefaultendpunct}{\mcitedefaultseppunct}\relax
\EndOfBibitem
\bibitem[Schwinger(1961)]{Schwinger1961}
Schwinger,~J. Brownian Motion of a Quantum Oscillator. \emph{J. Math. Phys.}
  \textbf{1961}, \emph{2}, 407--432\relax
\mciteBstWouldAddEndPuncttrue
\mciteSetBstMidEndSepPunct{\mcitedefaultmidpunct}
{\mcitedefaultendpunct}{\mcitedefaultseppunct}\relax
\EndOfBibitem
\bibitem[Keldysh(1964)]{Keldysh1964}
Keldysh,~L.~V. Diagram Technique for Nonequilibrium Processes. \emph{Zh. Eksp.
  Teor. Fiz.} \textbf{1964}, \emph{47}, 1018\relax
\mciteBstWouldAddEndPuncttrue
\mciteSetBstMidEndSepPunct{\mcitedefaultmidpunct}
{\mcitedefaultendpunct}{\mcitedefaultseppunct}\relax
\EndOfBibitem
\bibitem[{H.-P. Breuer and F. Petruccione}(2002)]{breuer}
{H.-P. Breuer and F. Petruccione}, \emph{Open Quantum Systems}; Oxford Univ.
  Press, 2002\relax
\mciteBstWouldAddEndPuncttrue
\mciteSetBstMidEndSepPunct{\mcitedefaultmidpunct}
{\mcitedefaultendpunct}{\mcitedefaultseppunct}\relax
\EndOfBibitem
\bibitem[Nitzan(2006)]{nitzan06}
Nitzan,~A. \emph{Chemical Dynamics in Condensed Phases: Relaxation, Transfer,
  and Reactions in Condensed Molecular Systems}; Oxford Univ. Press, 2006\relax
\mciteBstWouldAddEndPuncttrue
\mciteSetBstMidEndSepPunct{\mcitedefaultmidpunct}
{\mcitedefaultendpunct}{\mcitedefaultseppunct}\relax
\EndOfBibitem
\bibitem[Ryndyk and Cuniberti(2007)Ryndyk, and Cuniberti]{cuniberti2007}
Ryndyk,~B. S. D.~A.; Cuniberti,~G. Molecular Junctions in the Coulomb Blockade
  Regime: Rectification and Nesting. \emph{Phys. Rev. B} \textbf{2007},
  \emph{76}, 045408\relax
\mciteBstWouldAddEndPuncttrue
\mciteSetBstMidEndSepPunct{\mcitedefaultmidpunct}
{\mcitedefaultendpunct}{\mcitedefaultseppunct}\relax
\EndOfBibitem
\bibitem[Scheck(2013)]{Scheck2013}
Scheck,~F. \emph{Quantum Physics edn. 2}; Springer-Verlag Berlin Heidelberg,
  2013\relax
\mciteBstWouldAddEndPuncttrue
\mciteSetBstMidEndSepPunct{\mcitedefaultmidpunct}
{\mcitedefaultendpunct}{\mcitedefaultseppunct}\relax
\EndOfBibitem
\bibitem[Levy and Rabani(2013)Levy, and Rabani]{Levy2013}
Levy,~T.~J.; Rabani,~E. Symmetry Breaking and Restoration Using the
  Equation-of-Motion Technique for Nonequilibrium Quantum Impurity Models.
  \emph{J. Phys.: Condens. Matter} \textbf{2013}, \emph{25}, 115302\relax
\mciteBstWouldAddEndPuncttrue
\mciteSetBstMidEndSepPunct{\mcitedefaultmidpunct}
{\mcitedefaultendpunct}{\mcitedefaultseppunct}\relax
\EndOfBibitem
\bibitem[Levy and Rabani(2013)Levy, and Rabani]{Levy2013a}
Levy,~T.~J.; Rabani,~E. Steady State Conductance in a Double Quantum Dot Array:
  The Nonequilibrium Equation-of-Motion Green Function Approach. \emph{J. Chem.
  Phys.} \textbf{2013}, \emph{138}, 164125\relax
\mciteBstWouldAddEndPuncttrue
\mciteSetBstMidEndSepPunct{\mcitedefaultmidpunct}
{\mcitedefaultendpunct}{\mcitedefaultseppunct}\relax
\EndOfBibitem
\bibitem[Stefanucci \latin{et~al.}(2006)Stefanucci, Almbladh, Kurth, Gross,
  Rubio, van Leeuwen, Dahlen, and von Barth]{Stefanucci2006}
Stefanucci,~G.; Almbladh,~C.-O.; Kurth,~S.; Gross,~E. K.~U.; Rubio,~A.; van
  Leeuwen,~R.; Dahlen,~N.~E.; von Barth,~U. In \emph{Time-Dependent Density
  Functional Theory}; Marques,~M., Ullrich,~C., Nogueira,~F., Rubio,~A.,
  Burke,~K., Gross,~E., Eds.; Lecture Notes in Physics; Springer: Berlin, 2006;
  Vol. 706; p 479\relax
\mciteBstWouldAddEndPuncttrue
\mciteSetBstMidEndSepPunct{\mcitedefaultmidpunct}
{\mcitedefaultendpunct}{\mcitedefaultseppunct}\relax
\EndOfBibitem
\bibitem[Devreese and van Doren(1976)Devreese, and van Doren]{Langreth76}
Devreese,~J.~T., van Doren,~V.~E., Eds. \emph{Linear and Non-Linear Electron
  Transport in Solids}; Nato Science Series B; 17; Plenum: New York, 1976\relax
\mciteBstWouldAddEndPuncttrue
\mciteSetBstMidEndSepPunct{\mcitedefaultmidpunct}
{\mcitedefaultendpunct}{\mcitedefaultseppunct}\relax
\EndOfBibitem
\bibitem[Meir and Wingreen(1992)Meir, and Wingreen]{meir_landauer_1992}
Meir,~Y.; Wingreen,~N.~S. Landauer Formula for the Current Through an
  Interacting Electron Region. \emph{Phys. Rev. Lett.} \textbf{1992},
  \emph{68}, 2512--2515\relax
\mciteBstWouldAddEndPuncttrue
\mciteSetBstMidEndSepPunct{\mcitedefaultmidpunct}
{\mcitedefaultendpunct}{\mcitedefaultseppunct}\relax
\EndOfBibitem
\bibitem[Zubarev(1960)]{Zubarev1960}
Zubarev,~D.~N. Double-time Green Functions in Statistical Physics. \emph{Sov.
  Phys. Usp.} \textbf{1960}, 320–345\relax
\mciteBstWouldAddEndPuncttrue
\mciteSetBstMidEndSepPunct{\mcitedefaultmidpunct}
{\mcitedefaultendpunct}{\mcitedefaultseppunct}\relax
\EndOfBibitem
\bibitem[Bonch-Bruevich and Tiablikov(1962)Bonch-Bruevich, and
  Tiablikov]{Bruevich1962}
Bonch-Bruevich,~V.~L.; Tiablikov,~S.~V. \emph{The Green Function Method in
  Statistical Mechanics}; North-Holland Pub. Co.; Interscience Publishers
  Amsterdam, New York, 1962; p 251 p.\relax
\mciteBstWouldAddEndPunctfalse
\mciteSetBstMidEndSepPunct{\mcitedefaultmidpunct}
{}{\mcitedefaultseppunct}\relax
\EndOfBibitem
\end{mcitethebibliography}
\providecommand{\latin}[1]{#1}
\makeatletter
\providecommand{\doi}
  {\begingroup\let\do\@makeother\dospecials
  \catcode`\{=1 \catcode`\}=2\doi@aux}
\providecommand{\doi@aux}[1]{\endgroup\texttt{#1}}
\makeatother
\providecommand*\mcitethebibliography{\thebibliography}
\csname @ifundefined\endcsname{endmcitethebibliography}
  {\let\endmcitethebibliography\endthebibliography}{}

\end{document}